\documentclass[aps,prx,twocolumn,showpacs,floatfix,notitlepage]{revtex4-1}
\usepackage[latin9]{inputenc}
\usepackage{color}
\usepackage{verbatim}
\usepackage{units}
\usepackage{amsmath}
\usepackage{amssymb}
\usepackage{graphicx}
\usepackage{graphicx}  
\usepackage{epstopdf}
\usepackage{bm}        
\usepackage{amsmath}
\usepackage{amssymb}   
\usepackage{array}
\usepackage[caption=false]{subfig}
\usepackage[bookmarks=false,linkcolor=blue,urlcolor=blue,colorlinks,citecolor=blue]{hyperref}
\usepackage{dsfont}
\usepackage{appendix}

\makeatletter
\@ifundefined{textcolor}{}
{%
 \definecolor{BLACK}{gray}{0}
 \definecolor{WHITE}{gray}{1}
 \definecolor{RED}{rgb}{1,0,0}
 \definecolor{GREEN}{rgb}{0,1,0}
 \definecolor{BLUE}{rgb}{0,0,1}
 \definecolor{CYAN}{cmyk}{1,0,0,0}
 \definecolor{MAGENTA}{cmyk}{0,1,0,0}
 \definecolor{YELLOW}{cmyk}{0,0,1,0}
}

\usepackage{dcolumn}
\usepackage{bm}
\usepackage{setspace}
\usepackage{fancybox}
\usepackage{listings}
\usepackage{url}
\usepackage{pdfsync}
\usepackage{float}

\usepackage{subfig}
\begin{document}

\title{From an array of quantum wires to three-dimensional fractional topological insulators}

\author{Eran Sagi}

\author{Yuval Oreg}

\affiliation{Department of Condensed Matter Physics, Weizmann Institute of Science,
Rehovot, Israel 76100}
\begin{abstract}
The coupled-wires approach has been shown to be
useful in describing two-dimensional strongly interacting topological
phases. In this manuscript we extend this approach to three-dimensions,
and construct a model for a fractional strong topological insulator.
This topologically ordered phase has an exotic gapless
state on the surface, called a fractional Dirac liquid, which
cannot be described by the Dirac theory of free fermions. Like in non-interacting
strong topological insulators, the surface is protected by the presence
of time-reversal symmetry and charge conservation. We show that upon breaking these
symmetries, the gapped fractional Dirac liquid presents unique features.
In particular, the gapped phase that results from breaking time-reversal
symmetry has a halved fractional Hall conductance of the form $\sigma_{xy}=\frac{1}{2}\frac{e^{2}}{mh}$
 if the filling is $\nu=1/m$. On the other hand, if the surface is gapped by proximity
coupling to an $s$-wave superconductor, we end up with an exotic
topological superconductor. To reveal the topological nature of this
superconducting phase, we partition the surface into two regions:
one with broken time-reversal symmetry and another coupled to a superconductor.
We find a fractional Majorana mode, which cannot be described by a
free Majorana theory, on the boundary between the two regions. The
density of states associated with tunneling into this one-dimensional
channel is proportional to $\omega^{m-1}$, in analogy to the edge
of the corresponding Laughlin state.
\end{abstract}
\maketitle

\section{Introduction}

\label{sec:intro}

The theoretical predictions and consequent experimental discoveries
of topological insulators in two and three dimensions \cite{Pankratov1987,Kane2005,Bernevig2006a,Konig2007,Liu2012,Roth2009,Nowack2013,Knez2012,Spanton2014,Fu2007,Fu2007a,Hsieh2008}
have enriched our understanding of topological phases of matter. Importantly,
it was demonstrated that topological states may be much more widespread
than previously believed, and that in particular they can exist beyond
the realms of two-dimensions (2D). Since then, many systems realizing
various topological phases have been proposed, and a periodic table
for topological phases of non-interacting fermions has been established
\cite{Schnyder2008,Kitaev2009}. While this classification is limited
to topological states protected by time-reversal symmetry, particle-hole
symmetry, and chiral symmetry, the existing tools can be applied to
other protecting symmetries. A particularly notable extension of the
above classification are the topological crystalline insulators \cite{Fu2011},
protected by the crystal point group symmetries.

\begin{figure}
\includegraphics[scale=0.55]{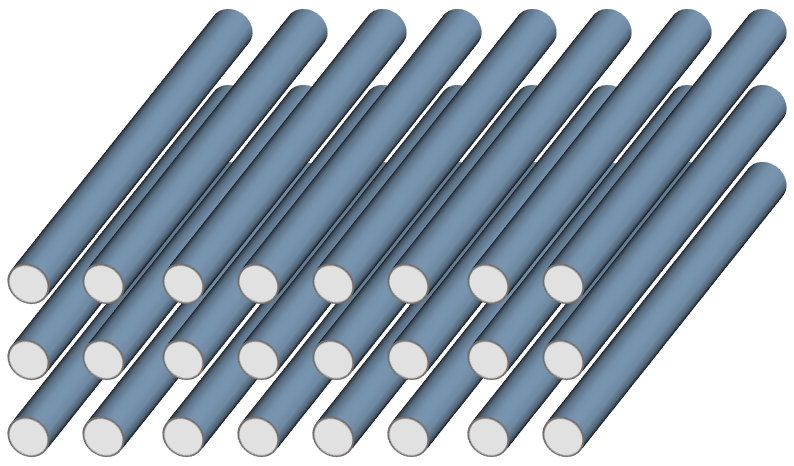}\protect\protect\caption{\label{fig:3D wires}A schematic depiction of the
array of wires studied in this manuscript. The coupled-wires approach
allows us to treat interacting terms using the Abelian bosonization
framework, making it very useful in constructing strongly interacting
topological phases. Throughout this work, we focus on the construction
of a fractional strong topological insulator. }
\end{figure}

The effects of strong interactions on topological materials, on the
other hand, are much more subtle. Particularly interesting situations
occur when interactions stabilize topologically ordered phases that
cannot be realized in free-fermion systems. The excitations in these
systems have fractional statistics, and usually carry fractional quantum
numbers. \textcolor{black}{Thus}, topologically ordered systems are
sometimes called fractional phases. The most famous example of such a phase
is the fractional quantum Hall effect (FQHE).

While topological phases have conclusively been found to exist in
three-dimensions (3D), fractional phases are still usually associated
with 2D systems. This is a consequence of the well-known theorem stating
that anyonic statistics between two point-particles cannot occur in
3D \cite{Doplicher1971,Doplicher1974}.

A possible way to go around this theorem and realize topologically
ordered phases in 3D is to consider loop excitations, which can have non-trivial braiding statistics with point-particles, as well as other loops.
\begin{figure*}
\textcolor{black}{}\subfloat[\label{fig:fractionalsxy}]{\textcolor{black}{\includegraphics[scale=0.6]{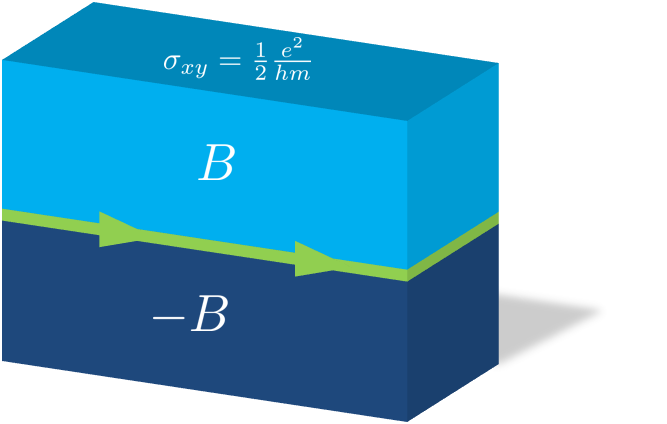}}}\textcolor{black}{}\subfloat[\label{fig:fractionalmajo}]{\textcolor{black}{\includegraphics[scale=0.6]{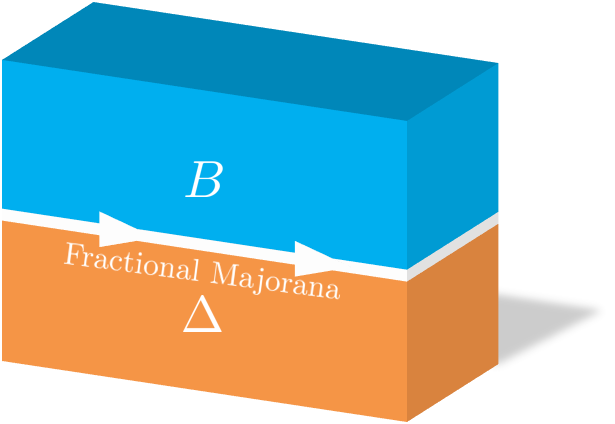}}}

\textcolor{black}{\protect\caption{The fractional strong topological insulator phase
we construct has an exotic fractional Dirac liquid on the surface.
As in the non-interacting case, this gapless surface is protected
by time reversal symmetry and charge conservation. Upon breaking any
of these symmetries, we end up with a gapped surface displaying unique
topological properties. For example, by breaking time-reversal symmetry,
we get a halved fractional quantum Hall effect, characterized by a
surface Hall conductance of the form $\sigma_{xy}=\frac{1}{2}\frac{e^{2}}{mh}$,
where $m$ is an odd integer (in a state associated with the 2D Laughlin
state at filling $\nu=1/m$). On the magnetic domain wall shown in
(a), one therefore finds a chiral Luttinger liquid, similar to the
edge mode of the corresponding 2D Laughlin-state. If the system is
gapped by proximity to an $s$-wave superconductor (i.e., by breaking
charge conservation), we get an exotic time-reversal invariant topological
superconductor. It is interesting to study the boundary between a
superconducting region and a magnetic region, as depicted in Fig (b).
A fractional Majorana mode, which cannot be described by a free Majorana
theory, is found on the interface between the two regions. The
tunneling density of states associated with this 1D channel is proportional
to $\omega^{m-1}$, in analogy to the edge of the corresponding Laughlin-state.}
}
\end{figure*}

Indeed, a few recent works \cite{Maciejko2010,Hoyos2010,Levin2011,Swingle2011,Maciejko2012,Swingle2012,Chan2013,Maciejko2014,Jian2014}
have used various non-perturbative approaches to discuss fractional
topological insulators in 3D. These are the fractional counterparts
of the well-studied \textcolor{black}{non-interacting} strong topological
insulators.

In this work we construct a model for such systems from an array of
weakly coupled wires, as illustrated in Fig. (\ref{fig:3D wires}). The ability to introduce interactions,
and treat them using the bosonization framework has made this approach
successful in producing fractional phases in 2D \cite{Kane2002,Teo2011,Klinovaja2013c,Neupert2014,Sagi2014,Klinovaja2014,Meng2014,Mross2014,Santos2015,Sagi2015,Gorohovsky2015,Meng2015}.
In addition to reproducing well-known FQHE states in the extremely
anisotropic limit, it has been shown to be useful in producing
new fractional states, which would otherwise be very hard to realize
in terms of a microscopic model.

Here we show that the wires approach can be extended to 3D, and use
it to construct a fractional strong topological insulator.

Since, by assumption, different wires are weakly coupled, we linearize
their energy spectra around the Fermi-points, and treat inter-wire
terms as perturbations within this linearized framework. Once the
1D spectra are linearized, we can fully describe the single and many
body excitations using chiral bosonic fields. The only remnants of
the original model are encoded in the values of the Fermi-momenta
associated with the various wires (these are determined by the overall
density, the spin-orbit coupling, the magnetic field, etc.). The Fermi-momenta
determine, due to momentum conservation, the allowed inter-wire terms.
To keep track of these terms, it proves useful to depict the Fermi-momenta
diagrammatically. In 2D models, for example, we plot the Fermi-momenta
as a function of an index enumerating the wires (see Figs. (\ref{fig:diagram_integer})-(\ref{fig:diagram_third})).
Importantly, the diagrams present the problem as a lattice model in
a lower dimension. The task of calculating some topological properties
of the full model is then reduced to the calculation of similar properties
in the lower-dimensional non-interacting system. For example, the
analysis of 2D topological insulators is reduced, in some aspects,
to the analysis of a 1D topological system. Similarly, the analysis
of a 3D strong topological insulator is reduced to the analysis of
a 2D topological insulator. At integer fillings, this enables us to
construct topological phases in two and three dimensions.

Upon varying the filling to a given fractional value, the pattern
formed by the Fermi-momenta is modified, and we are forced to consider
a different set of inter-wire terms, which in general need to involve
multi-electron processes. At a specific set of filling factors, we
find it useful to define new fields, associated with a new set of
effective Fermi-momenta. The transformation to the new fields can be chosen
such that the corresponding diagram is mapped onto the one describing
a system at integer filling. Then, by repeating the analysis that
led to the creation of the non-interacting topological phase in terms of the transformed
fields, we get its fractional analog.

The resulting 3D topologically ordered phase studied here will be
shown to have a novel gapless surface mode, which cannot be described
by Dirac's theory of free fermions. Throughout this work, we refer
to the surface as a fractional Dirac liquid. In analogy to a strong
topological insulator, the surface is protected by time reversal symmetry
and charge conservation.

It is insightful to study what happens to the fractional Dirac theory
once it is gapped by breaking any of these symmetries. We will see
that if a time reversal breaking perturbation is introduced, the surface
acquires a Hall conductance of the form $\sigma_{xy}=\frac{1}{2m}\frac{e^{2}}{h}$
in a state at filling $\nu=1/m$. This Hall conductance is half that
of the associated 2D Laughlin FQH state, and we therefore refer to
this effect as a halved fractional quantum Hall effect.
This should be compared with the half-integer quantum Hall effect
that occurs on the surface of a strong topological insulator, which
corresponds to $m=1$ in our framework. As a result of the above, a magnetic domain wall of the type depicted in Fig. (\ref{fig:fractionalsxy}) contains a gapless channel described by the chiral Luttinger-liquid theory, similar to the edge states of a Laughlin-state.

In addition, we will break charge conservation by proximity coupling
the surface to an $s$-wave superconductor. The resulting superconducting state is found to be topologically non-trivial. On the surface
of a non-interacting strong topological insulator, for example, one finds a phase
that resembles a $p_{x}+ip_{y}$ superconductor, but has time reversal
symmetry \cite{Fu2008}. One way to reveal the topological nature
of the surface is to separate the surface into two domains: one with
broken time-reversal symmetry, and another with broken charge conservation.
It was shown in Ref. \cite{Fu2008} that a chiral Majorana mode is
localized near the boundary separating the two regions. Furthermore, it was shown in Ref. \cite{Neupert2014a} that this remains correct in the presence of interactions.

Repeating the same thought experiment in our fractional phase, we
find a chiral self-Hermitian mode on the boundary, which cannot be
described by a free Majorana theory, as illustrated in Fig. (\ref{fig:fractionalmajo}). In particular, the tunneling
density of states associated with this 1D channel is proportional
to $\omega^{m-1}$, as opposed to the constant density of states characterizing
free Majorana fermions. Throughout this paper, we refer to this mode
as a fractional Majorana mode.

The structure of the paper is as follows: In Sec. \ref{sub:integer WTI}
we discuss the physics of fractional weak topological insulators.
These are states constructed by stacking 2D fractional topological
insulators \cite{Levin2009}. While these systems do not host anyonic
excitations which are deconfined in the stacking direction, they
possess interesting surface states, intimately related
to those of the fractional strong topological insulator. Then, in Sec.
\ref{sub:surface of strong}, we use an oversimplified yet intuitive
model (with a modified time-reversal symmetry), derived from the surface of the fractional weak topological
insulator, to describe the surface of the fractional strong topological
insulator. The purpose of this discussion is to pedagogically derive
the properties of the surface once it is gapped by breaking its protecting
symmetries. We stress that these properties will be derived rigorously
in Section \ref{sec:strong topological insulators} without using
the above oversimplified model.

In the sections that follow, we use
a wire construction to create a fractional strong topological insulator.
The analysis is done in a few stages. First, in Sec. \ref{sub:review of 2D}
we review the construction of a 2D fractional topological insulator
presented in Ref. \cite{Sagi2014}. This model will be the
starting point in our constructions of 3D states. In this section we also introduce
the general approach and notations used throughout the rest of this
work. Then, in Sec. \ref{sub:integer 3D}-\ref{sub:fractional 3D}
we will show that by stacking copies of this 2D phase with a set of
non-trivial inter-layer coupling terms, a fractional strong topological
insulator can be stabilized. Finally, in Sec. \ref{sub:Gapping the surface},
we will use the wires model to derive the properties of the resulting phase.

\section{Fractional weak topological insulators }

\subsection{Weak topological insulators}

\label{sub:integer WTI}

\begin{figure*}
\subfloat[\label{fig:stacking}]{\includegraphics[scale=0.53]{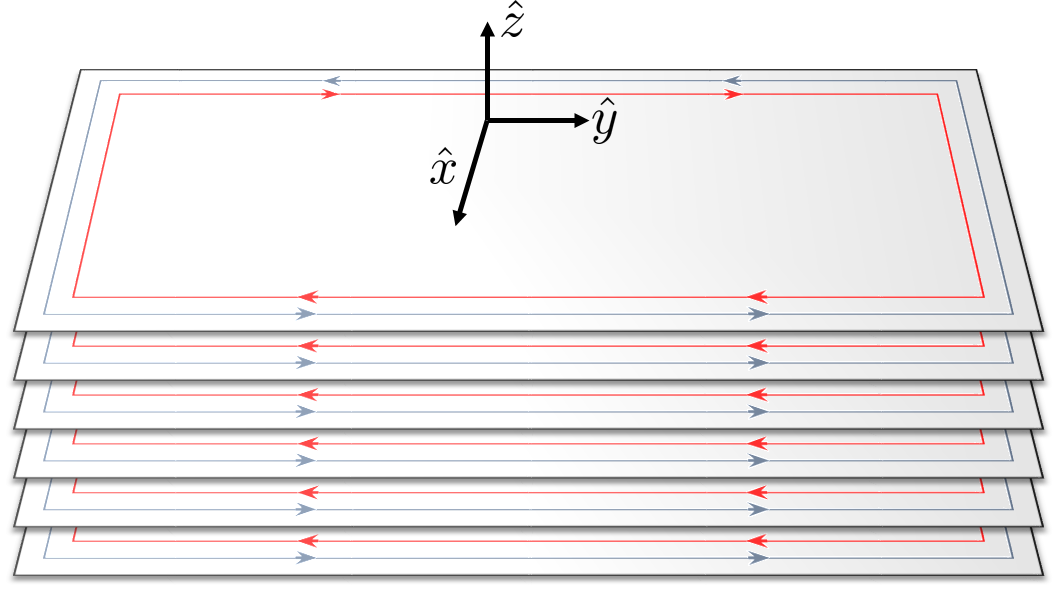}

}\subfloat[\label{fig:diagram_surface_coupling}]{\includegraphics[scale=0.4]{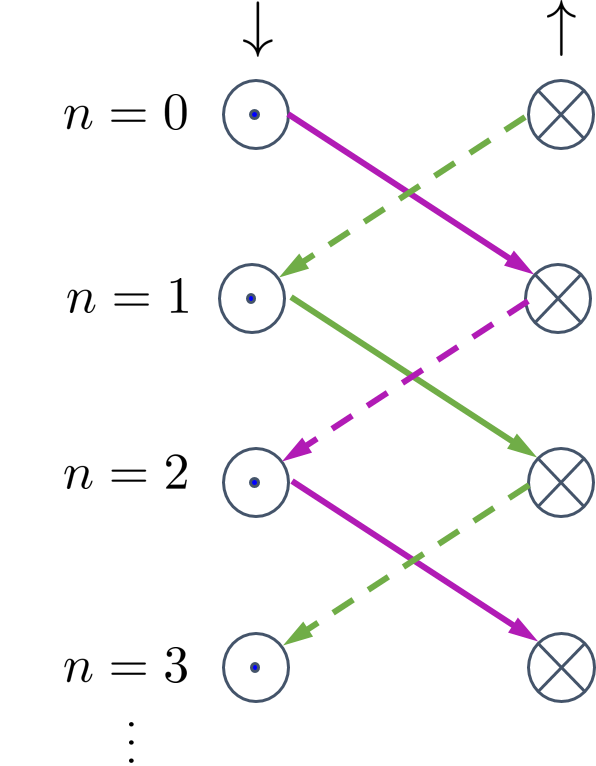}}

\protect\caption{(a) The simplified model we use to describe a weak topological insulator.
Each plane forms an $S_{z}$-conserving 2D topological insulator.
The edge of each layer therefore contains counter-propagating spin-up
and spin-down modes, represented by red and blue arrows. (b) A diagrammatic
depiction of the surface of the simplified model. The vertical direction
represents the layer index, and the horizontal direction represents
spin. The symbol $\bigotimes$ ($\bigodot$) corresponds to a right
(left) mover. We introduce nearest-layer coupling terms, represented
by arrows connecting different chiral modes. Time reversal symmetry
relates the amplitudes of the terms represented by the full arrows
to the amplitudes of the terms represented by the dashed arrows. The
above model can be decomposed into two decoupled gapless Hamiltonians,
represented by the green and purple arrows.}
\end{figure*}
Before discussing fractional weak topological insulators, we briefly
review the physics of non-interacting weak topological insulators.

To construct a simple model for a weak topological insulator, we imagine
stacking many 2D topological insulators (see Fig. (\ref{fig:stacking})
for illustration).

 For simplicity, we assume that the 2D topological insulators in the
various layers conserve $S_{z}$. We can therefore describe their
edge states as counter propagating spin-up and spin-down modes. In
the limit where the layers are decoupled, the surface in the $xz$
direction, for example, is composed of these helical modes. The modes
are represented diagrammatically in Fig. (\ref{fig:diagram_surface_coupling}),
where the vertical direction represents the layer index, and the horizontal
direction represents the spin. The symbol $\bigotimes$ ($\bigodot$)
corresponds to a right (left) mover.

We now introduce coupling between adjacent layers. In doing so, we
consider only the terms represented by the arrows in Fig. (\ref{fig:diagram_surface_coupling}),
as these are the only terms capable of gapping the surface. Time reversal
symmetry relates the amplitudes of the terms represented by the full
arrows and those represented by the dashed arrows.

To be concrete, if we define the electron annihilation operators $\psi_{n,\uparrow}(x)$,$\psi_{n,\downarrow}(x)$
(where $n$ is the layer index), we can write the low energy surface
Hamiltonian in the form
\begin{equation}
H=H_{z}+H_{x},\label{eq:H weak}
\end{equation}
where
\begin{align}
H_{x} & =-iv\sum_{n}\int dx\left(\psi_{n,\uparrow}^{\dagger}(x)\partial_{x}\psi_{n,\uparrow}(x)\right.\label{eq:Hx}\\
 & \left.\hphantom{fffff}-\psi_{n,\downarrow}^{\dagger}(x)\partial_{x}\psi_{n,\downarrow}(x)\right)\nonumber
\end{align}
represents the Hamiltonian of the decoupled helical edge modes, and
\begin{align}
H_{z} & =t\sum_{n}\int dx\left(\psi_{n,\uparrow}^{\dagger}(x)\psi_{n+1,\downarrow}(x)e^{i\alpha}\right.\label{eq:Hz}\\
 & \left.-\psi_{n,\downarrow}^{\dagger}(x)\psi_{n+1,\uparrow}(x)e^{-i\alpha}+h.c.\right)\nonumber
\end{align}
represents their nearest-layer coupling. In the above, $\alpha$ is
a fixed phase, which we set to be $0$ for simplicity.

If the system has periodic boundary conditions in the $x$ and $z$
directions, $k_{x}$ and $k_{z}$ are good quantum numbers. We can
therefore write the Hamiltonian in $k$-space, where it takes the
form $H=\sum_{k_{z}}\int dk_{x}\psi^{\dagger}(\vec{k})\mathcal{H}(\vec{k})\psi(\vec{k})$,
with $\psi(\vec{k})=(\begin{array}{cc}
\psi_{\uparrow}(\vec{k}), & \psi_{\downarrow}(\vec{k})\end{array})^{T}$ and
\begin{equation}
\mathcal{H}=-2t\sigma_{y}\sin\left(k_{z}a\right)+vk_{x}\sigma_{z}\label{eq:bloch hamiltonian}
\end{equation}
(here the $\sigma_{i}$'s are Pauli matrices acting on the spin degrees of freedom).
We see that we have two Dirac cones on the surface: one around $\left(k_{x}=0,k_{z}=0\right)$,
and another around $\left(k_{x}=0,k_{z}=\pi\right)$. These are not
protected by time-reversal symmetry, as time-reversal invariant terms
can couple the two Dirac modes and gap them out. However, such terms
necessarily involve a large momentum transfer. Therefore, if lattice
translation invariance is also imposed, the two Dirac modes remain
protected.

\subsection{The surface of fractional weak topological insulators}

\label{sub:fractional weak}

In order to generalize the above simplified model to a fractional
weak topological insulator, we imagine stacking layers of 2D fractional
topological insulators. Again, we focus on the $S_{z}$ conserving
case, where each 2D fractional topological insulator can be thought
of as two decoupled FQHE layers. These two FQH states have equal densities,
but opposite fillings $\pm\nu$ and spins.

For now, we keep the discussion general and do not provide a specific
model for the 2D fractional topological insulators in the various
layers. For such a concrete model, the reader is referred to Sec.
\ref{sub:review of 2D}, where we review the wire construction of
a fractional topological insulator introduced in Ref. \cite{Sagi2014}.

We focus on the time reversal invariant analog of a Laughlin state
with filling $\nu=1/m$, where $m$ is an odd integer. Before introducing
inter-layer coupling terms, the natural way to describe the edge modes
of the various layers is in terms of two boson fields $\chi_{n}^{\uparrow}\text{ and }\chi_{n}^{\downarrow}$.
These modes satisfy the commutation relations

\begin{widetext}
\begin{equation}
\left[\chi_{n}^{\sigma}(x),\chi_{n'}^{\sigma'}(x')\right]=\frac{1}{m}i\pi\sigma\delta_{n,n'}\delta_{\sigma,\sigma'}\text{{sign}}(x-x')+\frac{1}{m^{2}}i\pi\text{{sign}}(n-n')+\frac{1}{m^{2}}\pi\delta_{n,n'}\sigma_{y}^{\sigma,\sigma'},\label{eq:commutation of chi}
\end{equation}

\end{widetext}where in the above $\sigma=1(-1)$ for spin $\uparrow(\downarrow)$.
In terms of these, the chiral fermion operators take the form $\tilde{\psi}_{n,\sigma}\propto e^{im\chi_{n}^{\sigma}}$,
and the Laughlin quasiparticle operators are given by $\tilde{\psi}_{n,\sigma}^{QP}\propto e^{i\chi_{n}^{\sigma}}$.
If we introduce inter-layer coupling terms, the effective low
energy Hamiltonian describing the surface is given by
\begin{equation}
\tilde{H}=\tilde{H}_{z}+\tilde{H}_{x},\label{eq:H_tilde}
\end{equation}
where
\begin{equation}
\tilde{H}_{x}=\frac{mv}{4\pi}\sum_{n,\sigma}\int dx\left(\partial_{x}\chi_{n}^{\sigma}\right)^{2}\label{eq:Hx tilde}
\end{equation}
represents the decoupled chiral Luttinger liquids, and
\begin{align}
\tilde{H}_{z} & =\tilde{t}\sum_{n}\int dx\left(\tilde{\psi}_{n,\uparrow}^{\dagger}\tilde{\psi}_{n+1,\downarrow}\right.\label{eq:Hz tilde}\\
 & \left.\hphantom{jmk}-\tilde{\psi}_{n,\downarrow}^{\dagger}\tilde{\psi}_{n+1,\uparrow}+h.c.\right)\nonumber
\end{align}
represents their coupling. In what follows we assume that the amplitude
$\tilde{t}$ is large enough such that these operators flow to the strong coupling limit (if they are considered separately).
In the integer case, where $m=1$, $\tilde{H}$ reduces to the Hamiltonian $H$ defined in Eq. (\ref{eq:H weak}),
as can be seen directly by rewriting $H$ using Abelian bosonization.
In the fractional case, where $m>1$, $\tilde{H}$ cannot be mapped to a non-interacting fermionic Hamiltonian. However, the model can still be represented
by the diagram shown in Fig. (\ref{fig:diagram_surface_coupling}).
Notice that now the symbols $\bigotimes$ and $\bigodot$ represent right
and left moving $\tilde{\psi}$ (or $\chi$) fields, respectively.

Each of the two terms in Eq. (\ref{eq:Hz tilde}), if considered separately,
can gap out the spectrum in the thermodynamic limit. However, the
two gapped phases that result from these terms are topologically distinct.
Noting that the two non-commuting operators have the same amplitude
and scaling dimension, it is clear that the system must be in a critical
point between the two gapped phases. We emphasize that the criticality
of the surface is imposed by time reversal symmetry, and is not a result of fine-tuning
the Hamiltonian. It is explicitly assumed that additional interacting terms do not destabilize the critical point. This crucial assumption must be checked for a given microscopic model.

Like in the integer case, this gapless model is represented by two
decoupled gapless theories. This can be identified by writing the Hamiltonian
in the form $\tilde{H}=\tilde{H}_{1}+\tilde{H}_{2},$ with

\begin{widetext}

\begin{equation}
\tilde{H}_{1}=\frac{mv}{4\pi}\sum_{n,\sigma}\int dx\left(\partial_{x}\chi_{2n}^{\uparrow}\right)^{2}+\left(\partial_{x}\chi_{2n+1}^{\downarrow}\right)^{2}+\tilde{t}\sum_{n}\int dx\left(\tilde{\psi}_{2n,\uparrow}^{\dagger}\tilde{\psi}_{2n+1,\downarrow}-\tilde{\psi}_{2n+2,\uparrow}^{\dagger}\tilde{\psi}_{2n+1,\downarrow}+h.c.\right),\label{eq:H_1 tilde}
\end{equation}
\begin{equation}
\tilde{H}_{2}=\frac{mv}{4\pi}\sum_{n,\sigma}\int dx\left(\partial_{x}\chi_{2n+1}^{\uparrow}\right)^{2}+\left(\partial_{x}\chi_{2n}^{\downarrow}\right)^{2}+\tilde{t}\sum_{n}\int dx\left(\tilde{\psi}_{2n+2,\downarrow}^{\dagger}\tilde{\psi}_{2n+1,\uparrow}-\tilde{\psi}_{2n,\downarrow}^{\dagger}\tilde{\psi}_{2n+1,\uparrow}+h.c.\right).\label{eq:H_2 tilde}
\end{equation}

\end{widetext}

The green (purple) arrows in Fig. (\ref{fig:diagram_surface_coupling})
represent terms belonging to $\tilde{H}_{1}$ ($\tilde{H}_{2}$).
In the integer case each of these Hamiltonians is described by a low
energy Dirac theory, as can be seen by refermionizing the bosonic
theories. Importantly, the Hamiltonians $\tilde{H}_{1}$ and $\tilde{H}_{2}$
remain gapless in the fractional case as well, as indicated by the
argument we have used to show that $\tilde{H}$ is gapless. However,
in the fractional case these Hamiltonians cannot be described by a
low energy Dirac theory. In the future we refer to their low energy
theories as fractional Dirac liquids.

\section{A simplified model for the surface of fractional strong topological
insulators}

\label{sub:surface of strong}

\begin{figure*}
\subfloat[\label{fig:AFTI-1}]{\includegraphics[scale=0.53]{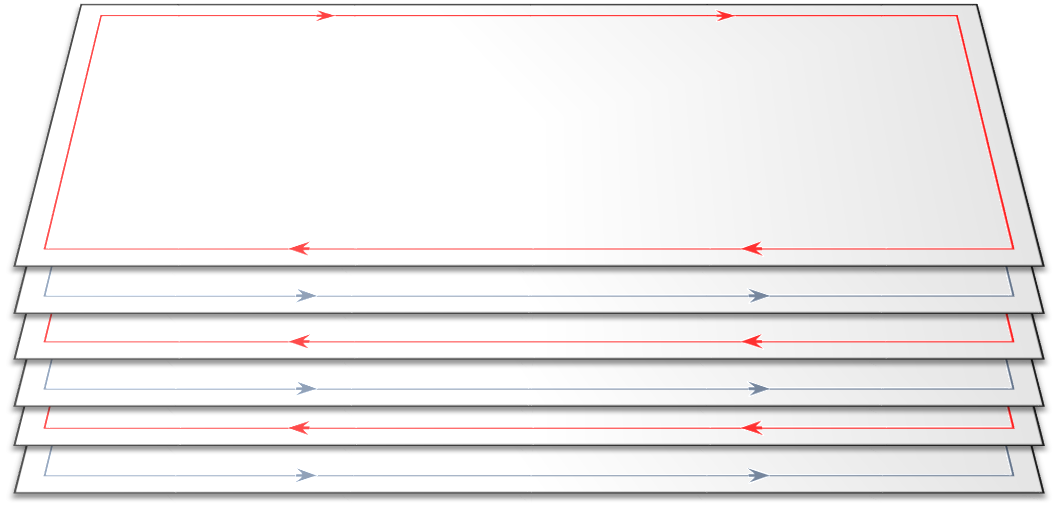}

}\subfloat[\label{fig:AFTI-DIAGRAM}]{\includegraphics[scale=0.4]{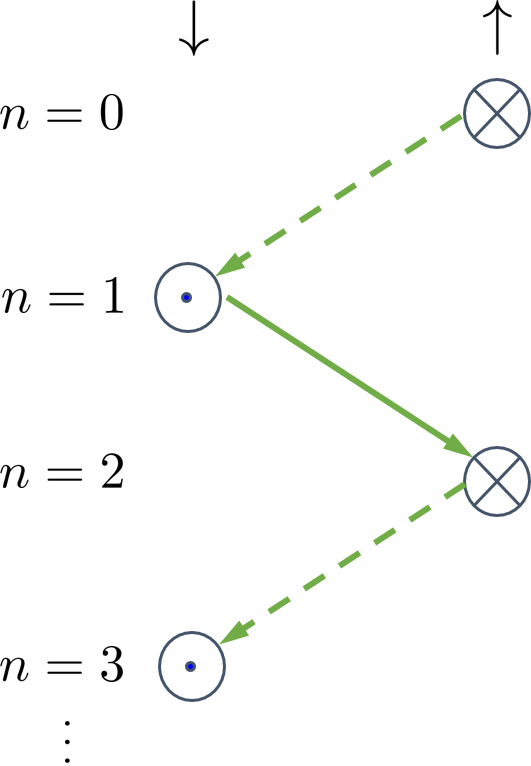}}\subfloat[\label{fig:AFTI diagram TR breaking}]{\includegraphics[scale=0.4]{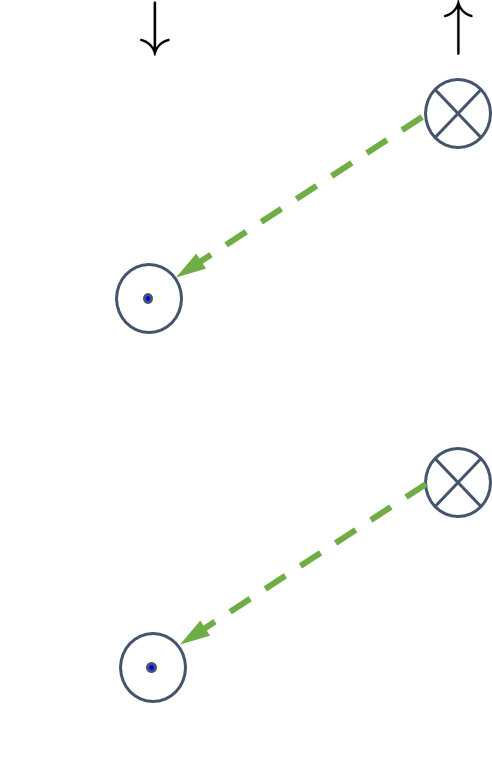}

}\protect\caption{(a) The simplified model we use to describe an antiferromagnetic topological
insulator and its fractional analog. Each layer can be thought of as a single-component
QHE system at filling $\pm\nu$ (where $\nu=1$ in the integer case, and
in the fractional case we consider $\nu=1/m$, where $m$ is an odd
integer). Due to the existence of a modified time-reversal symmetry,
given by the product of the original time-reversal operator and a
translation by a unit cell, the $xz$ surface remains gapless even
in the presence of inter-layer coupling terms. The surface can be
described in terms of coupled 1D channels, which enables a simple
derivation of universal surface properties. We argue that these properties
remain true for a strong topological insulator and its fractional
analog, where the surface is protected by the local time reversal
symmetry. (b) A diagrammatic representation of $xz$ surface of the
above model in terms of the coupled 1D chiral edge modes. (c) The
diagram that corresponds to the situation where the modified time-reversal
symmetry is explicitly broken by introducing a term of the form (\ref{eq:breaking time reversal invariance}) with $\tilde{t}'=\tilde{t}$. }
\end{figure*}

We have seen above that the simple model for a fractional weak topological
insulator can be decomposed into two decoupled gapless theories described
by the Hamiltonians $\tilde{H}_{1}$ and $\tilde{H}_{2}$. In the
integer case, weak topological insulators possess an even number of Dirac cones on the surface, while strong topological
insulators necessarily have an odd number of Dirac cones \cite{Fu2007}. By analogy,
we naturally expect the low energy theory describing the surface of
a fractional strong topological insulator to be related to the surface
of a fractional weak topological insulator. We therefore ask whether
a single decoupled surface Hamiltonian, say $\tilde{H}_{1}$, describing
left movers on layers with even indices and right movers on layers
with odd indices (see Fig. (\ref{fig:AFTI-1})), can faithfully describe
the surface of a fractional strong topological insulator.

Strictly speaking, this is clearly impossible since $\tilde{H}_{1}$
is not independently time reversal invariant. Nevertheless, we follow
Ref. \cite{Mross2014} in noting that $\tilde{H}_{1}$ is invariant
under a modified time reversal operation, defined as the product of
the original time-reversal operator and a translation by a unit cell.

Such a symmetry characterizes antiferromagnetic topological insulators,
which can be created, for example, by adding an antiferromagnetic
order parameter to a strong topological insulator (without closing
the gap). Remarkably, it was shown in Refs. \cite{Mong2010,Fang2013}
that the introduction of such a time-reversal breaking perturbation
does not destroy all the surface Dirac cones. Instead, the remaining
Dirac cones are protected by the modified time-reversal operation
described above.

This leads us to assume that $\tilde{H}_{1}$ faithfully describes
the surface of the fractional analog of an antiferromagnetic topological
insulator. The corresponding gapless surface is expected to be in
the same universality class as the surface of a fractional strong
topological insulator with a local time-reversal symmetry. We can
therefore use $\tilde{H}_{1}$ to derive some of the universal properties
expected to characterize the surface of a fractional strong topological
insulator. Notice that the bulk excitations of the fractional antiferromagnetic topological insulator are generally different from the fractional excitations characterizing the 3D fractional strong topological insulator.

As noted in Ref. \cite{Mross2014}, studying a system with the modified time-reversal symmetry,  for which one can write an explicit model of the surface in terms of weakly coupled 1D channels, greatly simplifies the analysis. This is impossible in systems that have a local (unmodified) time-reversal symmetry. To study these using a set of weakly coupled 1D systems, we will be forced to explicitly construct the 3D bulk as well. This will be done in Sec. \ref{sec:strong topological insulators}.

Again, we depict the surface model using diagrams representing the
inter-layer terms coupling the various chiral modes. In particular,
the diagram that represents the gapless surface Hamiltonian $\tilde{H}_{1}$
is shown in Fig. (\ref{fig:AFTI-DIAGRAM}).

Until now we have preserved the modified time-reversal symmetry and
charge conservation of $\tilde{H}_{1}$, and thus the surface was
found to be gapless. In what follows we focus on the properties of
the surface once these protecting symmetries are explicitly broken.
We will see below that the resulting gapped fractional Dirac mode
has unique properties.

To break the modified time reversal symmetry, we add a perturbation
of the form

\begin{align}
\tilde{H}_{t} & =\tilde{t}'\sum_{n}\int dx\left(\tilde{\psi}_{2n,\uparrow}^{\dagger}\tilde{\psi}_{2n+1,\downarrow}\right.\nonumber \\
 & \left.\hphantom{jmk}+\tilde{\psi}_{2n+2,\uparrow}^{\dagger}\tilde{\psi}_{2n+1,\downarrow}+h.c.\right)\label{eq:breaking time reversal invariance}
\end{align}
The physics of the resulting phase becomes transparent at the points
$\tilde{t}'=\pm\tilde{t}$, where the Hamiltonian is decomposed into
decoupled sine-Gordon models. For example, if $\tilde{t}'=\tilde{t}$
the non-quadratic part of the Hamiltonian takes the form
\begin{align}
2\tilde{t}'\sum\int dx\left(\tilde{\psi}_{2n,\uparrow}^{\dagger}\tilde{\psi}_{2n+1,\downarrow}+h.c.\right)=\nonumber \\
4\tilde{t}'\sum\int dx\cos\left[m\left(\chi_{2n}^{\uparrow}-\chi_{2n+1}^{\downarrow}\right)\right].\label{eq:t'=00003Dt}
\end{align}
Thus we end up with a set of mutually commuting cosine terms, which
can be pinned in the ground state and fully gap the surface.  This configuration is depicted
diagrammatically in Fig. (\ref{fig:AFTI diagram TR breaking}).
\begin{figure*}
\includegraphics[scale=0.6]{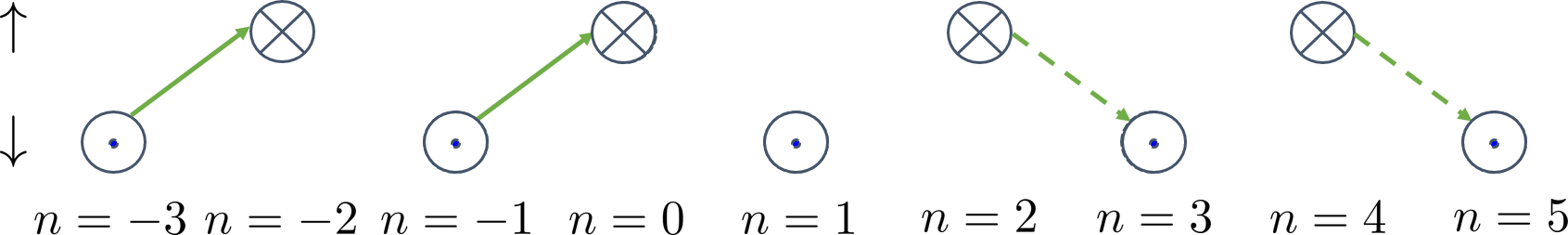}

\protect\caption{\label{fig:AFTI_TR_DOMAIN} A diagrammatic representation, in terms
of the $\tilde{\psi}$ (or $\chi$) modes defined at the beginning
of Sec. \ref{sub:fractional weak}, of the situation where the coefficient
$\tilde{t}'$ of the time reversal breaking term changes abruptly
from $-\tilde{t}$ to $\tilde{t}$ around $n=1$. A decoupled chiral $\tilde{\psi}$
mode is seen to be localized near the boundary at $n=1$, leading
to the conclusion that the gapped surface has a halved fractional
quantum Hall effect.}
\end{figure*}

To identify the surface Hall conductance $\sigma_{xy}$ characterizing
this gapped state, we change $\tilde{t}'$ abruptly from $-\tilde{t}$
to $\tilde{t}$ around the point $n=1$, as demonstrated in Fig. (\ref{fig:AFTI_TR_DOMAIN}).
It can be seen that a chiral $\chi$-mode, localized on the boundary,
remains decoupled. This mode is identical to the mode residing on
the edge of a $\nu=1/m$ Laughlin FQHE state. By invoking
the bulk-edge correspondence and noticing that this mode is a result
of contributions from the two sides of the boundary, we find that
the gapped surface has $\sigma_{xy}=\pm\frac{1}{2m}\frac{e^{2}}{h}.$ We note that since the perturbation defined in Eq. (\ref{eq:breaking time reversal invariance}) is small compared to the bulk energy gap, the resulting gapless chiral mode and the associated value of $\sigma_{xy}$ must be interpreted as properties of the surface.

Thus, the gapped surface of the fractional strong topological insulator
exhibits a halved fractional quantum Hall effect. We emphasize that
the above result is not limited to the special choice of parameters, \textcolor{black}{which were tuned to the so called ``sweet point''
\cite{Kitaev2001,Seroussi2014}, }and is in fact true as long as the
gap remains open. Physically, deviations from $\tilde{t}'=\pm\tilde{t}$
in the two regions introduce a non-zero localization length for the
boundary mode.

Alternatively, the fractional Dirac liquid can be gapped out by proximity-coupling
to an $s$-wave superconductor, i.e., by breaking charge conservation.
A superconducting term which does not violate the modified time-reversal
symmetry is given by
\begin{align}
\tilde{H}_{\varDelta} & =\Delta\sum_{n}\int dx\left[\tilde{\psi}_{2n,\uparrow}\tilde{\psi}_{2n+1,\downarrow}\right.\nonumber \\
 & \left.\phantom{}\phantom{}+\tilde{\psi}_{2n,\uparrow}\tilde{\psi}_{2n-1,\downarrow}+h.c.\right].\label{eq:proximity coupling}
\end{align}

\begin{figure*}
\subfloat[\label{fig:time reversal breaking with Majorana}]{\includegraphics[scale=0.6]{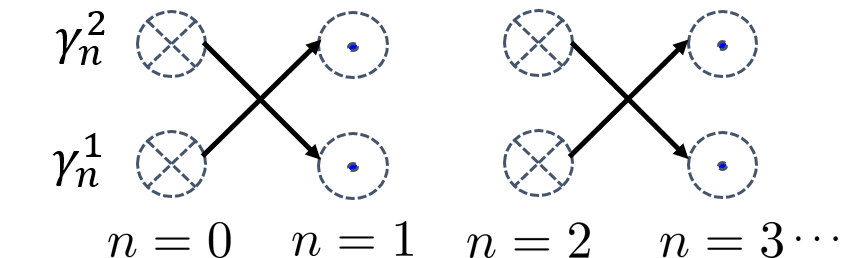}}\subfloat[\label{fig:AFTI_SC}]{\includegraphics[scale=0.6]{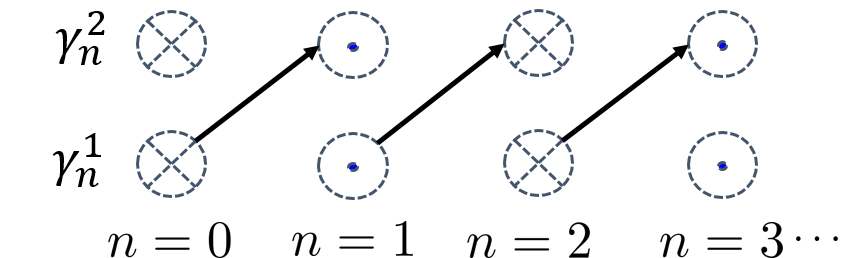}}

\subfloat[\label{fig:AFTI_TR_SC_DOMAIN}]{\includegraphics[scale=0.6]{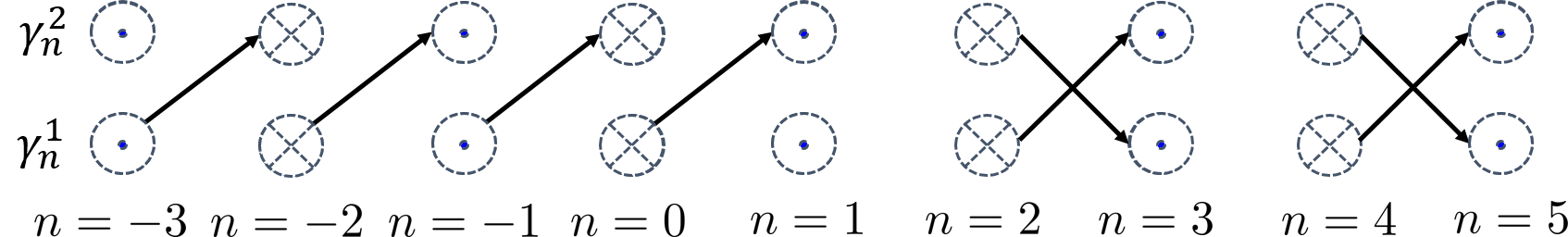}}

\protect\caption{A diagrammatic representation of the surface model in terms of the
fractional Majorana modes $\gamma_{n}^{i}$ (defined after Eq. (\ref{eq:proximity in terms of Majorana}))
in three situations: (a) The modified time reversal symmetry is broken
by the perturbation (\ref{eq:breaking time reversal invariance})
with $\tilde{t}'=\tilde{t}$. (b) Charge conservation is broken by
a superconducting term of the form (\ref{eq:proximity coupling})
with $\Delta=\tilde{t}$. (c) The region with $n<1$ of the surface is
gapped by a superconducting term, and the region with $n>1$ is gapped
by a modified time-reversal symmetry breaking term. A chiral fractional
Majorana mode is attached to the boundary at $n=1$. Here the vertical
direction represents the index $i$, and the horizontal direction
represents the layer index $n$. Dotted symbols represent the chiral
fractional Majorana modes.}
\end{figure*}

Again, the physics becomes simple at a sweet point given by $\Delta=\pm\tilde{t}$.
Taking $\Delta=\tilde{t}$, for example, the non-quadratic part of the Hamiltonian
can be written as
\begin{equation}
\tilde{H}_{\varDelta}=2i\tilde{t}\sum_{n}\left(-1\right)^{n}\int dx\gamma_{n}^{1}\gamma_{n+1}^{2},\label{eq:proximity in terms of Majorana}
\end{equation}
with $\tilde{\psi}_{n,\sigma}=\gamma_{n}^{1}+i\gamma_{n}^{2},$ and $\left(\gamma_{n}^{i}\right)^{\dagger}=\gamma_{n}^{i}$
(notice that we have omitted the spin index, which is fully determined
by $n$). The modes $\gamma_{n}^{i}$ are referred to as fractional
Majorana modes. Indeed, in the integer case, these become chiral Majorana
modes, described by a free Majorana theory. The structure of this
Hamiltonian is depicted diagrammatically in Fig. (\ref{fig:AFTI_SC}),
where the dotted symbols represent the fractional Majorana modes $\gamma_{n}^{1}$
and $\gamma_{n}^{2}$. From the form of Eq. (\ref{eq:proximity in terms of Majorana}),
it is clear that the system is gapped. Notice that Eq. (\ref{eq:breaking time reversal invariance}),
describing time-reversal symmetry breaking, can be expressed in terms
of the fractional Majorana modes as well. In particular, the non-quadratic
part of the Hamiltonian $\tilde{H}_{1}+\tilde{H}_{t}$ with $\tilde{t}'=\tilde{t}$
takes the simple form $2i\tilde{t}\sum_{n}\int dx\left(\gamma_{2n+1}^{1}\gamma_{2n}^{2}-\gamma_{2n+1}^{2}\gamma_{2n}^{1}\right)$, which is depicted in Fig. (\ref{fig:time reversal breaking with Majorana}).

Once the surface is gapped by proximity to a superconductor, it forms
an exotic time-reversal invariant topological superconductor. We leave
the further investigation of such a novel superconducting state to
future works.

However, it is illuminating to examine the boundary
between a magnetic region with broken time-reversal symmetry, and
a superconducting region. The physics is simplest if we gap the region with
$n>1$ using a time reversal breaking term of the form (\ref{eq:breaking time reversal invariance})
with $\tilde{t}'=\tilde{t}$, and the region with $n<1$ using a proximity-coupling
term of the form (\ref{eq:proximity coupling}) with $\Delta=\tilde{t}$.
This situation is depicted in Fig. (\ref{fig:AFTI_TR_SC_DOMAIN}).
We see that a decoupled chiral Majorana mode of the form $\gamma_{1}^{1}=\frac{1}{2}\left(e^{im\chi_{1}^{\downarrow}}+e^{-im\chi_{1}^{\downarrow}}\right)$
is localized on the boundary. The propagator describing this fractional
Majorana field takes the form $\left\langle \gamma_{1}^{1}(x,t)\gamma_{1}^{1}(0,0)\right\rangle \propto(x+vt)^{-m}$,
making it clear that this gapless channel cannot be described by a
free Majorana theory. We point out the similarity to the edge mode
of the 2D fractional topological superconductor discussed in Ref.~\cite{Vaezi2013}.

One can generalize the construction to study hierarchical states, with $\nu$ different from $1/m$. Within the simplified model, each layer in Fig. (\ref{fig:AFTI-1}) now contains a FQHE state with a general $K$-matrix of dimension $d$, and therefore $d$ distinct edge modes $\chi_n^{j}$ (where $n$ is the layer index and $j=1,\cdots,d$ enumerates the edge modes in a given layer). Switching on terms that preserve the modified time reversal symmetry and couple only modes with the same $j$, we get a generalized fractional Dirac theory associated with any $K$-matrix. It is clear, however, that the generalized fractional Dirac theories are not necessarily protected by symmetries, as some modes can be gapped out by coupling fields with different $j$'s (without breaking time-reversal symmetry). 

If we repeat the analysis presented in this section and break the modified time-reversal symmetry (still without coupling modes with different $j$'s), we get a halved fractional Hall conductance of the form $\sigma_{xy}=\frac{\nu e^2}{2h}$. Additional coupling between modes with different $j$'s has the potential of changing the Hall conductance. Therefore, if symmetry breaking perturbations are introduced, a number of topologically distinct gapped phases may arise on the surface.

The above analysis relied on a modified time reversal symmetry to
directly model the surface using coupled 1D modes. In what follows
we treat a 3D model with a local time reversal symmetry. As we will
see, the analysis presented in the next sections reproduces the universal
results derived here.

\section{Fractional Strong topological insulators}
\label{sec:strong topological insulators}
\subsection{Construction of 2D fractional topological insulators}
\label{sub:review of 2D}
In Sec. \ref{sub:integer 3D}-\ref{sub:fractional 3D}
we will construct a 3D model for a fractional strong topological insulator.
Our starting point will be the 2D construction of a fractional topological
insulator which was introduced in Ref. \cite{Sagi2014}. In this section
we review this construction, and introduce the general approach used throughout the rest of this work. Unlike the pervious section (Sec.~\ref{sub:surface of strong}), where we used a simplified model with a modified time-reversal symmetry [see Fig.~(\ref{fig:AFTI-1})], here we have a local (unmodified) time reversal symmetry. To avoid confusion, we will use a different set of notations in the analysis that follows.

\subsubsection{Two-dimensional topological insulator from an array of quantum wires}

\begin{figure}
\includegraphics[scale=0.6]{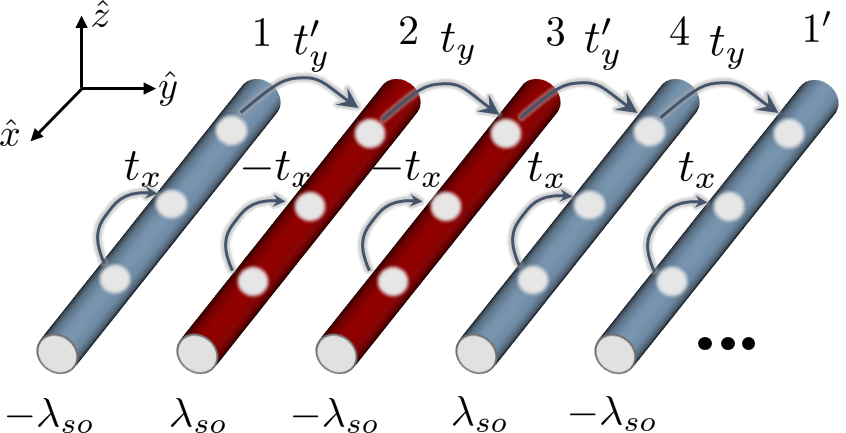}\protect\caption{\label{fig:wires}The wire construction we use as a realization of
a 2D fractional topological insulator. This model is the starting
point in our construction of a 3D fractional strong topological insulator
 (see Sec. \ref{sub:integer 3D}-\ref{sub:fractional 3D}). }
\end{figure}

The 2D model which is the focus of this section is made of $4N$ wires, as
depicted in Fig. (\ref{fig:wires}). The $j$'s unit cell is composed
of four wires. We note that for convenience the unit cell is shifted
by one wire with respect to the unit cell defined in Ref. \cite{Sagi2014}.

 The first and last wires in each unit cell contain electrons, i.e.,
their highest occupied states are close to the minimum of the conduction
band. In a particle-hole symmetric fashion, the two other wires contain
holes, i.e., their highest occupied states are close to the maximum
of the band.

We describe the system in terms of a tight binding model, where each
wire is composed of sites at positions $x=a\cdot i$ (see Fig. (\ref{fig:wires})).
Here $a$ is the distance between adjacent sites, and $i$ is an integer
enumerating the sites. Adjacent lattice points are coupled with a
hopping amplitude $t_{x}$ ($-t_{x}$) in the electron (hole) wires,
as depicted by the arrows in Fig. (\ref{fig:wires}). We define the
annihilation operator $\psi_{\alpha,s}^{\left(i,j\right)}$ for an
electron with spin $s$ in wire number $\alpha$ ($\alpha=1\ldots4$)
of the unit cell labeled by $\left(i,j\right)$. The Hamiltonian that
describes the decoupled wires takes the form $H_{0}=H_{m}+H_{x}$,
where
\begin{align}
H_{m} & =m\sum_{i,j}\left(\sum_{\alpha=1,4}\psi_{\alpha,s}^{(i,j)\dagger}\psi_{\alpha,s}^{(i,j)}-\right.\label{eq:H_m}\\
 & \left.-\sum_{\alpha=2,3}\psi_{\alpha,s}^{(i,j)\dagger}\psi_{\alpha,s}^{(i,j)}+h.c.\right),\nonumber
\end{align}
and
\begin{align}
H_{x} & =-t_{x}\sum_{i,j}\left(\sum_{\alpha=1,4}\psi_{\alpha,s}^{(i+1,j)\dagger}\psi_{\alpha,s}^{(i,j)}-\right.\label{eq:H_x}\\
 & \left.-\sum_{\alpha=2,3}\psi_{\alpha,s}^{(i+1,j)\dagger}\psi_{\alpha,s}^{(i,j)}+h.c.\right).\nonumber
\end{align}

The term $H_{x}$ describes intra-wire hopping, and the term $H_{m}$
produces an opposite shift in energy for the electrons and the holes.
We note that to have a particle-hole symmetry, the chemical potential
is set to be zero.
Going to $k$-space, we write $H=\sum_{\mathbf{k}\in BZ1}\vec{\psi}{}^{\dagger}(\mathbf{k})\mathcal{H}(\mathbf{k})\vec{\psi}(\mathbf{k}),$
with $\vec{\psi}=\left(\begin{array}{cccccccc}
\psi_{1\uparrow} & \psi_{2\uparrow} & \psi_{3\uparrow} & \psi_{4\uparrow} & \psi_{1\downarrow} & \psi_{2\downarrow} & \psi_{3\downarrow} & \psi_{4\downarrow}\end{array}\right)^{T}$. In addition, we define the matrices \textbf{$\tau_{i}$ }($i=1\ldots3$)\textbf{
}as the Pauli-matrices operating on the wires 1-2
(and 3-4) space. Similarly, the matrices $\sigma_{i}$
operate on the (1,2)-(3,4) blocks, and $s_{i}$
act on the spin degrees of freedom.

In terms of these, the Bloch Hamiltonian corresponding to Eqs. (\ref{eq:H_m})-(\ref{eq:H_x})
takes the form
\begin{equation}
\mathcal{H}_{0}=\left[m-2t_{x}\cos\left(k_{x}a\right)\right]\sigma_{z}\tau_{z}.\label{eq:h0}
\end{equation}

\begin{figure*}
\subfloat[\label{fig:spectrum_integer}]{\includegraphics[scale=0.35]{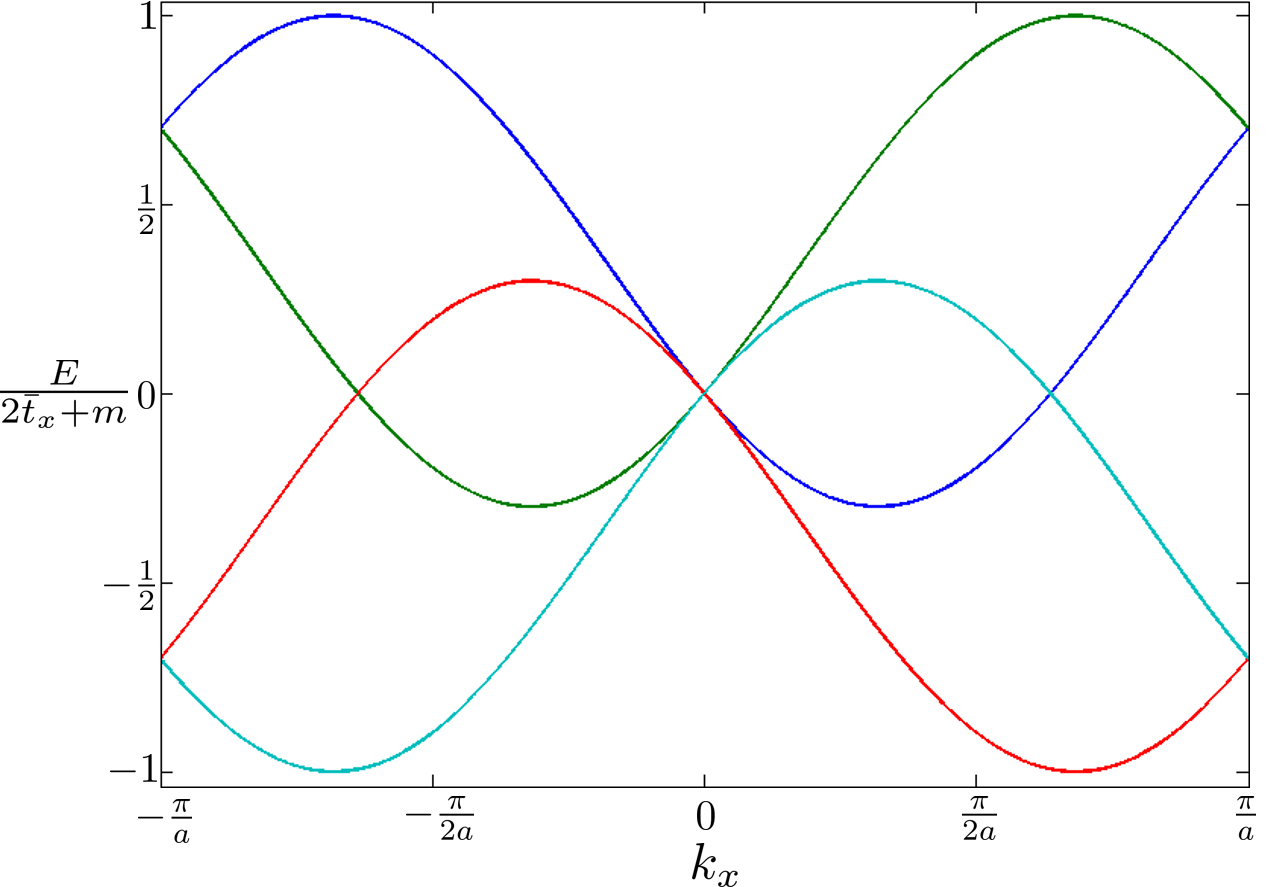}

}\subfloat[\label{fig:spectrum_fractional}]{\includegraphics[scale=0.35]{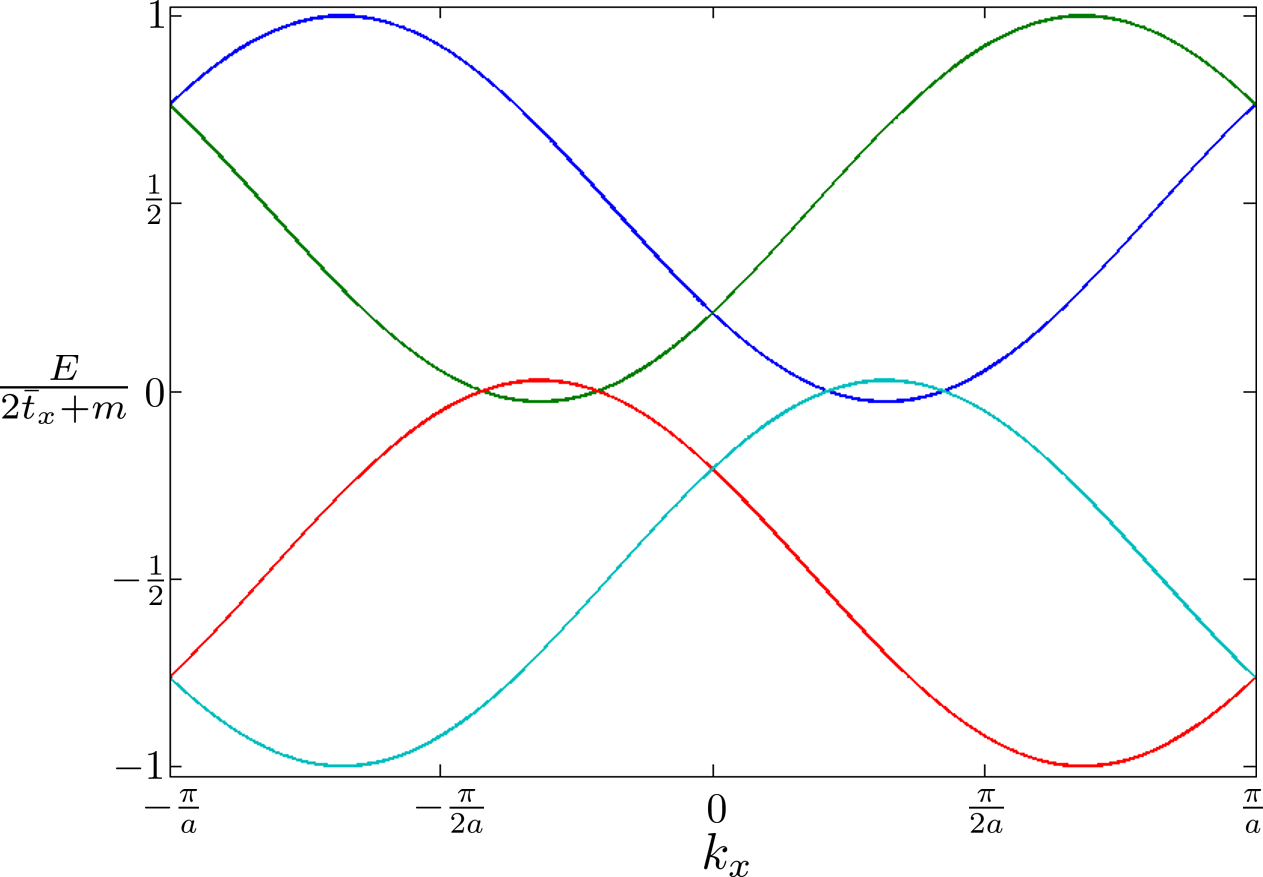}}

\protect\caption{The spectrum of the system depicted in Fig. (\ref{fig:wires}), when
the inter-wire terms are switched off for (a) $\nu=1$, and (b) $\nu=1/3$.
We note that only the spin-up sector is presented here. The energies
in blue, cyan, red, and green correspond to wires number 1, 2, 3,
and 4, respectively. }
\end{figure*}

Furthermore, we introduce Rashba spin-orbit interactions with an alternating
coupling $\lambda_{\text{{so}}}\left(-1\right)^{\alpha}$. If the
electric field is aligned in the $y$ direction, the resulting term
is
\begin{equation}
\mathcal{H}_{\text{so}}=-\lambda_{\text{{so}}}s_{z}\tau_{z}\sin\left(k_{x}a\right).\label{eq:hso}
\end{equation}
It proves convenient to define new parameters: $\bar{t}_{x},k_{\text{so}},\text{ and }k_{F}^{0},$
satisfying
\begin{align}
t_{x} & =\bar{t}_{x}\cos\left(k_{\text{so}}a\right),\label{eq:relabeling}\\
\lambda_{\text{so}} & =2\bar{t}_{x}\sin\left(k_{\text{so}}a\right),\nonumber \\
m & =2\bar{t}_{x}\cos\left(k_{F}^{0}a\right).\nonumber
\end{align}
In terms of these, it is simple to see that the energy spectra of
the four decoupled wires in each unit cell take the form
\begin{align}
E_{1,s} & =2\bar{t}_{x}\left[\cos\left(k_{F}^{0}a\right)-\cos\left(\left(k_{x}-sk_{\text{so}}\right)a\right)\right],\nonumber \\
E_{2,s} & =-2\bar{t}_{x}\left[\cos\left(k_{F}^{0}a\right)-\cos\left(\left(k_{x}-sk_{\text{so}}\right)a\right)\right],\nonumber \\
E_{3,s} & =-2\bar{t}_{x}\left[\cos\left(k_{F}^{0}a\right)-\cos\left(\left(k_{x}+sk_{\text{so}}\right)a\right)\right],\nonumber \\
E_{4,s} & =2\bar{t}_{x}\left[\cos\left(k_{F}^{0}a\right)-\cos\left(\left(k_{x}+sk_{\text{so}}\right)a\right)\right].\label{eq:energies without tunneling}
\end{align}
If we define the filling factor as
\begin{equation}
\nu=\frac{k_{F}^{0}}{k_{\text{so}}},\label{eq:filling factor}
\end{equation}
the spectra corresponding to the spin-up sector are depicted in Fig.
(\ref{fig:spectrum_integer}) for the $\nu=1$ case, and in Fig. (\ref{fig:spectrum_fractional})
for the $\nu=1/3$ case.
We now introduce small tunneling operators that couple adjacent wires,
and write the inter-wire Hamiltonian in the form $H_{y}+H_{y}^{'}$,
where
\begin{align}
H_{y} & =-t_{y}\sum_{i,j}\left[\psi_{3,s}^{(i,j)\dagger}\psi_{2,s}^{(i,j)}+\psi_{1,s}^{(i,j+1)\dagger}\psi_{4,s}^{(i,j)}+h.c.\right],\label{eq:H_y}\\
H_{y}^{'} & =-t_{y}'\sum_{i,j}\left[\psi_{2,s}^{(i,j)\dagger}\psi_{1,s}^{(i,j)}+\psi_{4,s}^{(i,j)\dagger}\psi_{3,s}^{(i,j)}+h.c.\right],\label{eq:Hy'}
\end{align}
with $t_{y},t'_{y}\ll\overline{t}_{x}$. The
parameter $t_{y}$ describes hopping between two electron-wires
or two hole-wires, whereas $t'_{y}$ couples the electron and hole
wires.

The Bloch Hamiltonian can now be written in the form

\begin{widetext}
\begin{align}
\mathcal{H} & =2\overline{t}_{x}\left[\left(\cos\left(k_{F}^{0}a\right)-\cos\left(k_{x}a\right)\cos\left(k_{so}a\right)\right)\sigma_{z}\tau_{z}-s_{z}\tau_{z}\sin\left(k_{so}a\right)\sin\left(k_{x}a\right)\right]\nonumber \\
 & \phantom{}\phantom{}-t'_{y}\tau_{x}-\frac{t_{y}}{2}\left(\tau_{y}\sigma_{y}+\tau_{x}\sigma_{x}\right)-\frac{t_{y}}{2}\left(\tau_{x}\sigma_{x}-\tau_{y}\sigma_{y}\right)\cos\left(4k_{y}a\right)-\frac{t_{y}}{2}\left(\tau_{x}\sigma_{y}+\tau_{y}\sigma_{x}\right)\sin\left(4k_{y}a\right).\label{eq:h}
\end{align}

\end{widetext}

For simplicity, we treat the case where the lattice spacings are identical
in the two directions: $a_{x}=a_{y}\equiv a$. This requirement can
be lifted without affecting any of the topological properties. Notice
that in these conventions, the first Brillouin zone is defined as\textbf{
$k_{y}\in\left(-\frac{\pi}{4a},\frac{\pi}{4a}\right],k_{x}\in\left(-\frac{\pi}{a},\frac{\pi}{a}\right].$ }
We first investigate the integer case, $\nu=1$, where it can be checked
from Eq. (\ref{eq:h}) that as long as $t_{y}\neq t_{y}'$ and $t_{y}'\neq0$,
the system is completely gapped when periodic boundary conditions
are employed. At $t_{y}=t_{y}'$ the gap closes, indicating that there
might be a phase transition between two topologically distinct phases.
To understand the nature of the insulating phases for different values
of $t_{y}\text{ and }t'_{y}$, it is useful to linearize the spectrum
around the Fermi-momenta.

Assuming that $t_{y},t_{y}'\ll t_{x}$, the inter-wire terms can be
treated as perturbations within the linearized theory. Once the spectrum
has been linearized, the only remnants of the original model are the
Fermi momenta of the various modes. If momentum conservation is imposed,
the values of the Fermi momenta severely restrict the allowed inter-wire
coupling terms for a given $\nu$. To keep track of these terms, it
is convenient to present the Fermi-momenta as a function of the wire
index ($i,\alpha$), as depicted in Fig. (\ref{fig:diagram_integer})
for the spin-up sector.

Again, the symbol $\bigotimes$ ($\bigodot$) represents a right
(left) moving mode, and the arrows represent the coupling between different
modes, generated by the terms defined in Eqs. (\ref{eq:H_y})-(\ref{eq:Hy'}).
Since the system is fully gapped for $0<t'_{y}<t_{y}$, any such state
is adiabatically connected, and therefore topologically equivalent,
to the state where $t'_{y}$ is negligibly small compared to $t_{y}$.
Therefore, it is simple to see that the terms $\psi_{2,\uparrow}^{i,R\dagger}\psi_{3,\uparrow}^{i,L}+h.c.$
and $\psi_{4\uparrow}^{i,R\dagger}\psi_{1,\uparrow}^{i+1,L}+h.c.$
(and in the same way, $\psi_{2,\downarrow}^{i,L\dagger}\psi_{3,\downarrow}^{i,R}+h.c.$
and $\psi_{4\downarrow}^{i,L\dagger}\psi_{1,\downarrow}^{i+1,R}+h.c.$
for the spin-down sector) dominate, and gap the modes near $k_{x}=0$.
These terms, however, leave two gapless modes on each edge of the
system - one for each spin. For the edge at $i=1$, these edge modes
are $\psi_{1,\uparrow}^{1,L},\text{ and }\psi_{1,\downarrow}^{1,R}$.
The terms $\psi_{1\uparrow}^{i,R\dagger}\psi_{2,\uparrow}^{i,L}+h.c.$
and $\psi_{3\uparrow}^{i,R\dagger}\psi_{4,\uparrow}^{i,L}+h.c.$ (and
in the same way $\psi_{1\downarrow}^{i,L\dagger}\psi_{2,\downarrow}^{i,R}+h.c.$
and $\psi_{3\downarrow}^{i,L\dagger}\psi_{4,\downarrow}^{iR}+h.c.$)
are responsible for gapping the modes with $k_{x}\neq0$. We therefore
have a fully gapped bulk with counter-propagating edge modes, protected
by time reversal symmetry. Thus, the phase with $t'_{y}<t_{y}$ is
a topological insulator.

On the other hand, any state with $t'_{y}>t_{y}$ is adiabatically
connected to the state with $t_{y}=0$, where now the terms $\psi_{1,\uparrow}^{i,L\dagger}\psi_{2,\uparrow}^{i,R}+h.c.$
and $\psi_{3,\uparrow}^{i,L\dagger}\psi_{4,\uparrow}^{i,R}+h.c$.
(and in the same way $\psi_{1,\downarrow}^{i,R\dagger}\psi_{2,\downarrow}^{i,L}+h.c.$
and $\psi_{3,\downarrow}^{i,R\dagger}\psi_{4,\downarrow}^{i,L}+h.c$)
gap the modes near $k_{x}=0$. In this case we have a fully gapped
bulk with no edge modes, indicating that the phase we discuss is topologically
trivial.

\subsubsection{Two-dimensional fractional topological insulator from an array of
quantum wires}

Having found that the above model (defined in Eqs. (\ref{eq:H_m})-(\ref{eq:H_x}) and (\ref{eq:H_y})-(\ref{eq:Hy'}))
can be a topological insulator in the $\nu=1$ case, we now turn to
study fractional fillings. We restrict ourselves here to fillings
of the form $\nu=1/m$, where $m$ is an odd integer. More general
situations were discussed in Ref. \cite{Sagi2014}.

\begin{figure}
\subfloat[\label{fig:diagram_integer}]{\includegraphics[scale=0.33]{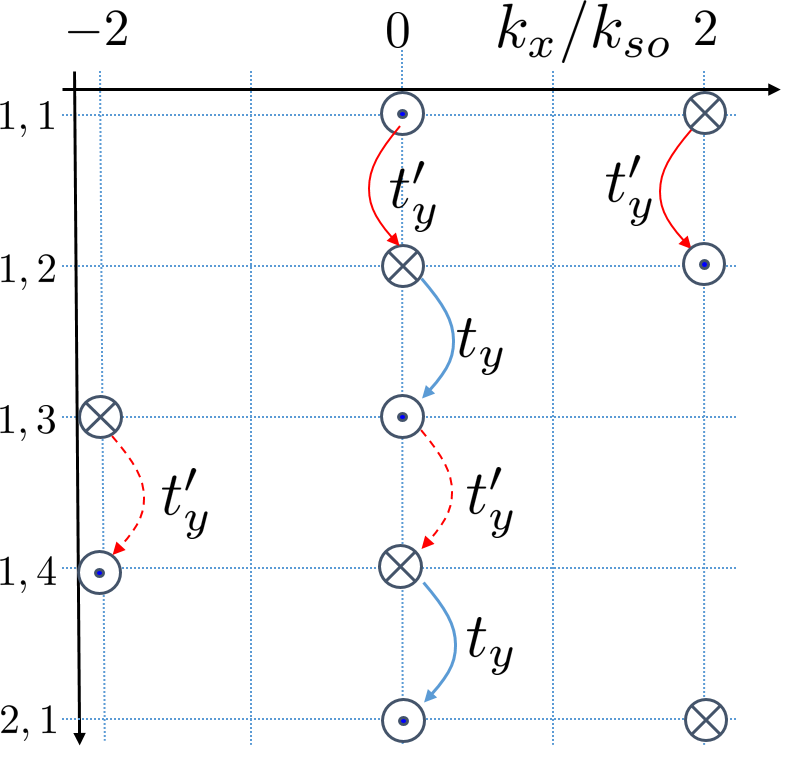}}
\subfloat[\label{fig:diagram_third}]{\includegraphics[scale=0.33]{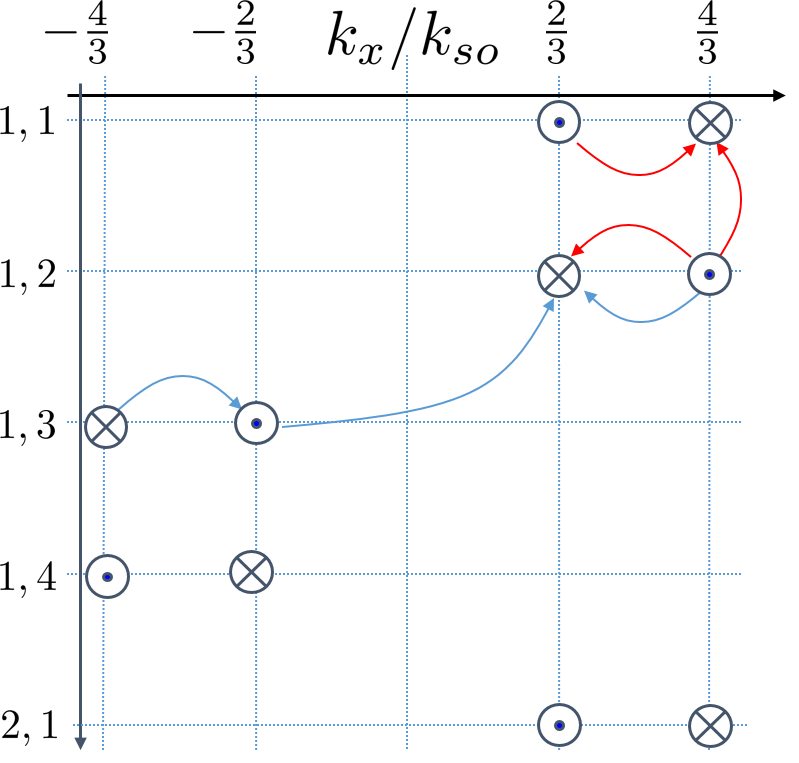}}\protect\caption{The diagrams which present the Fermi-momenta of the 2D model as a
function of the wire index (labeled by $(i,\alpha)$,
where $i$ and $\alpha$ are indices enumerating the unit cells and
the position within the unit cell, respectively) for (a) $\nu=1$
, and (b) $\nu=1/3$. Notice that only the modes with spin-up are
presented here.}
\end{figure}

The diagram that corresponds to the $\nu=1/3$ case is shown in Fig.
(\ref{fig:diagram_third}). It is evident that simple tunneling processes
do not conserve momentum and are therefore incapable of gapping the
system in this case. Since we are interested in a gapped phase, we
have to introduce interactions and consider multi-electron processes.

Fortunately, interaction terms become manageable if we use the standard
Abelian bosonization technique. We define the chiral boson fields
$\phi_{\alpha,s}^{i,R/L}$, such that $\psi_{\alpha,s}^{i,R/L}\propto\exp\left[i\left(\phi_{\alpha,s}^{i,R/L}+k_{\alpha,s}^{i,R/L}x\right)\right]$,
where $k_{\alpha,s}^{i,R/L}$ is the corresponding Fermi-momentum,
and the boson fields satisfy the commutation relations
\begin{widetext}
\begin{align}
\left[\phi_{\alpha,s}^{n,\rho},\phi_{\alpha',s'}^{n',\rho'}\right] & =i\rho\pi\delta_{s,s'}\delta_{\alpha,\alpha'}\delta_{\rho,\rho'}\delta_{n,n'}\text{sign}(x-x')\label{eq:commutations of phis}\\
 & \hphantom{d}+i\pi\left(\text{sign}(n-n')+\delta_{n,n'}\text{sign}(\alpha-\alpha')\right)+\delta_{n,n'}\delta_{\alpha,\alpha'}\pi\left(\sigma_{y}^{s,s'}+\delta_{s,s'}\sigma_{y}^{\rho,\rho'}\right),\nonumber
\end{align}
\end{widetext}
ensuring that the electron operators anti-commute. It is useful to
define new chiral fermion operators according to
\begin{align}
\tilde{\psi}_{\alpha,s}^{i,R/L} & =\left[\psi_{\alpha,s}^{i,R/L}\right]^{\frac{m+1}{2}}\left[\left(\psi_{\alpha,s}^{i,L/R}\right)^{\dagger}\right]^{\frac{m-1}{2}}\label{eq:psi tilde}\\
 & \propto e^{i\left(\eta_{\alpha,s}^{i,R/L}+q_{\alpha,s}^{i,R/L}x\right)},\nonumber
\end{align}
with
\begin{align}
\eta_{\alpha,s}^{i,R/L} & =\frac{m+1}{2}\phi_{\alpha,s}^{i,R/L}-\frac{m-1}{2}\phi_{\alpha,s}^{i,L/R},\label{eq:eta}\\
q_{\alpha,s}^{i,R/L} & =\frac{m+1}{2}k_{\alpha,s}^{i,R/L}-\frac{m-1}{2}k_{\alpha,s}^{i,L/R}.\label{eq:q}
\end{align}
We note that Eq. (\ref{eq:psi tilde}) makes sense only when the\textbf{
}Fermionic operators of the same type are located at close but separate
points in space. The commutation relations of the new bosonic $\eta$-fields
are

\begin{widetext}
\begin{align}
\left[\eta_{\alpha,s}^{n,\rho},\eta_{\alpha',s'}^{n',\rho'}\right] & =i\rho m\pi\delta_{s,s'}\delta_{\alpha,\alpha'}\delta_{\rho,\rho'}\delta_{n,n'}\text{sign}(x-x')\label{eq:commutation of etas}\\
 & \hphantom{d}+i\pi\left(\text{sign}(n-n')+\delta_{n,n'}\text{sign}(\alpha-\alpha')\right)+\delta_{n,n'}\delta_{\alpha,\alpha'}\pi\left(\sigma_{y}^{s,s'}+m\delta_{s,s'}\sigma_{y}^{\rho,\rho'}\right).\nonumber
\end{align}

\end{widetext}

Plotting the diagrams corresponding to the $q$'s, we find that the
picture is identical to the one associated with $\nu=1$ (Fig. (\ref{fig:diagram_integer})).
The transformation from the original Fermionic degrees of freedom
to the composite $\tilde{\psi}$-fields, which takes a simple linear
form in terms of the boson fields, can therefore be interpreted as
a transformation from $\nu=1/m$ to $\nu=1$. Hence, in terms of the
$\tilde{\psi}$-fields, we can repeat the analysis of the $\nu=1$
case, and write the terms used to obtain a topological insulator.
Writing these terms using the $\eta$ bosonic fields, we have:
\begin{align}
H_{t} & =\sum_{i}\left[\int dx\tilde{t}_{y}\cos\left(\eta_{4,\uparrow}^{i,R}-\eta_{1,\uparrow}^{i+1,L}\right)+R\longleftrightarrow L,\uparrow\longleftrightarrow\downarrow\right.\nonumber \\
 & +\int dx\tilde{t}_{y}\cos\left(\eta_{2,\uparrow}^{i,R}-\eta_{3,\uparrow}^{i,L}\right)+R\longleftrightarrow L,\uparrow\longleftrightarrow\downarrow\nonumber \\
 & +\sum_{s}\sum_{\rho=R,L}\int dx\tilde{t'}_{y}\cos\left(\eta_{1,s}^{i,\rho}-\eta_{2,s}^{i,\overline{\rho}}\right)\nonumber \\
 & \left.+\sum_{s}\sum_{\rho=R,L}\int dx\tilde{t}'_{y}\cos\left(\eta_{3,s}^{i,\rho}-\eta_{4,s}^{i,\overline{\rho}}\right)\right].\label{eq:eta terms}
\end{align}

Notice that while these are simple tunneling operators in terms of
the $\tilde{\psi}$-fields, they describe multi-electron processes
of the type depicted by the arrows in Fig. (\ref{fig:diagram_third}),
in terms of the original $\psi$-fields. It was shown in Refs. \cite{Kane2002,Teo2011,Oreg2013}
that there is a large range of strong density-density interactions
for which such terms flow to the strong coupling limit. Furthermore,
for all density-density interactions, the terms $\tilde{t'}_{y}$
and $\tilde{t}_{y}$ flow to the strong coupling limit if their bare values are
large enough, as their flow diagram coincides with that of the Sine-Gordon
model.

Similar to the integer case, if $\tilde{t}'_{y}<\tilde{t}_{y}$ the
first two terms in Eq. (\ref{eq:eta terms}) dominate. Assuming they
are made relevant, it is evident that the bulk becomes gapped and
that the two modes $\eta_{1,\uparrow}^{i,L}\text{ and }\eta_{1,\downarrow}^{i,R}$
remain decoupled. Each of these modes is similar to the edge mode
of a $\nu=1/m$ Laughlin QHE state. One can now follow Ref. \cite{Teo2011}
and show that the bulk excitations are fractionally charged and have fractional
statistics.

The above analysis suggests that the fractional analog of a given
integer topological phase can be realized if we manage to construct
the integer phase from an array of coupled wires. Once this is done, we can obtain a
fractional phase by considering an analogous system where the $\psi$-fields
are replaced by the $\tilde{\psi}$-fields (or equivalently,
$\phi$ is replaced by $\eta$), as demonstrated above.

This motivates us to use a similar approach in the construction of
fractional phases in 3D. Toward this goal, in the next section we
will construct a non-interacting strong topological insulator from an array of wires.
Then, in Sec. \ref{sub:fractional 3D} we will study the fractional
phase obtained by reducing the filling to $\nu=1/m$, and constructing a strong topological insulator
in terms of the composite $\tilde{\psi}$-fields.

\subsection{Strong topological insulators from weakly coupled wires}

\label{sub:integer 3D}
In order to construct a strong topological insulator we stack 2D layers,
each made of the wire construction presented in Sec. \ref{sub:review of 2D},
and tune the system such that $\nu=1$. The resulting 3D system is
made of an array of wires, as illustrated in Fig. (\ref{fig:3D wires}).
For simplicity, we assume that the distance between adjacent layers
is $a$ as well. The goal of this section is to engineer a time-reversal invariant
system with a single Dirac cone near the first and last layers.

To do so, we start by tuning each layer to the critical point between
the topological and the trivial phase, such that the 2D bulk contains
two Dirac modes. This can be achieved by taking $t_{y}=t'_{y}$, in
which case the Bloch Hamiltonian describing a single layer is given
by

\begin{widetext}
\begin{align}
\mathcal{H} & _{xy}=2\overline{t}_{x}\left[\left(\cos\left(k_{\text{so}}a\right)-\cos\left(k_{x}a\right)\cos\left(k_{\text{so}}a\right)\right)\sigma_{z}\tau_{z}-s_{z}\tau_{z}\sin\left(k_{\text{so}}a\right)\sin\left(k_{x}a\right)\right]\nonumber \\
 & \phantom{}\phantom{}\phantom{}\phantom{}-t_{y}\tau_{x}-\frac{t_{y}}{2}\left(\tau_{y}\sigma_{y}+\tau_{x}\sigma_{x}\right)-\frac{t_{y}}{2}\left(\tau_{x}\sigma_{x}-\tau_{y}\sigma_{y}\right)\cos\left(4k_{y}a\right)-\frac{t_{y}}{2}\left(\tau_{x}\sigma_{y}+\tau_{y}\sigma_{x}\right)\sin\left(4k_{y}a\right).\label{eq:hxy}
\end{align}

\end{widetext}To create a topologically non-trivial gapped 3D phase,
we perturb the above gapless Hamiltonian by an inter-layer term of
the form

\begin{align}
\mathcal{H}_{z} & =\frac{1}{2}\left[\left(m'-2t_{z}\cos\left(k_{z}a\right)\right)\tau_{x}\right.\label{eq:hz}\\
 & \hphantom{ddddd}\left.+2t'_{z}\sin\left(k_{z}a\right)s_{y}\tau_{z}\right]\left(1-\sigma_{x}\right),\nonumber
\end{align}
with $m',t_{z},t'_{z}\ll t_{y}\ll\bar{t}_{x}$. Unless
otherwise noted, we implicitly assume that all the coupling constants
are positive.

To see under which circumstances this model forms a strong topological
insulator, we now show that if the system is cut at the $z=0$ plane,
a single Dirac mode is localized near the resulting surface. It is clear that
as long as the inter-layer coupling terms are small, the important
degrees of freedom are those close to the Dirac points in each layer.
We therefore project the Hamiltonian onto the low-energy subspace
of the intra-layer Hamiltonian (Eq. (\ref{eq:hxy})).

The two Dirac cones are located at $k_{x}=k_{y}=0$. Therefore, to
identify this low-energy subspace we solve the equation
\begin{equation}
\mathcal{H}_{xy}(k_{x}=k_{y}=0)\psi_{0}=0\label{eq:low energy subspace}
\end{equation}
for the vectors $\psi_{0}$. The resulting subspace can be spanned
by the four vectors $\left|\pm1,s\right\rangle \equiv\left|\sigma_{x}=-1,\tau_{x}=\pm1,s\right\rangle $,
defined such that $\sigma_{x}\left|\pm1,s\right\rangle =-\left|\pm1,s\right\rangle ,\tau_{x}\left|\pm1,s\right\rangle =\pm\left|\pm1,s\right\rangle ,S_{z}\left|\pm1,s\right\rangle =s\left|\pm1,s\right\rangle $.
In what follows, the basis vectors are ordered as $\{\left|1,\uparrow\right\rangle ,\left|-1,\downarrow\right\rangle ,\left|1,\downarrow\right\rangle \left|-1,\uparrow\right\rangle \}$.
As expected, the $xy$ Hamiltonian, projected onto the low-energy
subspace and expanded to first order in momenta, takes the form of
two Dirac cones. To be specific, in the above basis the intra-layer Hamiltonian takes
the form

\begin{equation}
\mathcal{H}_{xy}=\left(\begin{array}{cc}
0 & h_{\text{Dirac}}\\
h_{\text{Dirac}}^{\dagger} & 0
\end{array}\right),\label{eq:hxy in critical point}
\end{equation}

where

\begin{widetext}
\begin{equation}
h_{\text{Dirac}}=\left(\begin{array}{cc}
0 & -2\overline{t}_{x}\sin\left(k_{\text{so}}a\right)k_{x}a+2it_{y}k_{y}a\\
2\overline{t}_{x}\sin\left(k_{\text{so}}a\right)k_{x}a-2it_{y}k_{y}a & 0
\end{array}\right).\label{eq:hDirac}
\end{equation}

\end{widetext}

Once the inter-layer part of the Hamiltonian is projected onto the same low energy subspace, it takes the form
\begin{equation}
\mathcal{H}_{z}=\left(\begin{array}{cc}
h_{\text{1D}}(k_{z}) & 0\\
0 & h_{\text{1D}}^{*}(k_{z})
\end{array}\right),\label{eq:hz(h1d)}
\end{equation}
where $h_{\text{1D}}(k_{z})=\vec{d}\cdot\vec{\sigma}$,
with $\vec{d}=\left(0,2t'_{z}\sin\left(k_{z}a\right),m'-2t_{z}\cos\left(k_{z}a\right)\right)$, can be thought of as the Hamiltonian a 1D model.

If $m'<2t_{z}$, the planar vector $\vec{d}$ winds once around the
origin as $k_{z}$ winds around the Brillouin zone, indicating that
the 1D model is topologically non-trivial. Indeed, in this regime
the model can be shown to have zero-energy end modes.

Since the $\hat{z}$ part of our 3D Hamiltonian consists of the two
decoupled models $h_{1D}$ and $h_{1D}^{*}$, it produces two zero-energy
modes on the $z=0$ surface. It is clear that the full low energy
Hamiltonian, projected onto the subspace spanned by these two
end modes, must be that of a single Dirac cone.

To see this explicitly, we need an analytic form for the two end modes.
This becomes simple if we restrict ourselves to the regime where $2t_{z}$
is close to $m$ and the gap is small. In this limit, focusing again
on the low energy properties, we can expand Eq. (\ref{eq:hz(h1d)})
in small $k_{z}$, and get a continuum model. The model $h_{1D}$
then takes the form
\begin{equation}
h_{1D}\approx\left(\begin{array}{cc}
m'-2t_{z}-t_{z}\partial_{z}^{2} & -2t'_{z}a\partial_{z}\\
2t'_{z}a\partial_{z} & -\left(m'-2t_{z}-t_{z}\partial_{z}^{2}\right)
\end{array}\right)\label{eq:real space h1D}
\end{equation}
in the position basis.

Assuming that the system ends at $z=0$, we look for localized zero-energy
eigenstates of this Hamiltonian and its complex conjugate, which vanish at $z=0$. Plugging
in an exponentially decaying function as an ansatz, we find two such
solutions
\begin{equation}
\left|1\right\rangle =\frac{1}{\sqrt{2}}\left(\begin{array}{c}
1\\
1\\
0\\
0
\end{array}\right)f(z),\left|2\right\rangle = \frac{1}{\sqrt{2}}\left(\begin{array}{c}
0\\
0\\
1\\
-1
\end{array}\right)f(z),\label{eq:explicit form of end modes of h1d}
\end{equation}
with $f(z)=\frac{\sqrt{2\lambda_{+}\lambda_{-}\left(\lambda_{+}+\lambda_{-}\right)}}{\lambda_{+}-\lambda_{-}}\left(e^{-\lambda_{+}z}-e^{-\lambda_{-}z}\right)$,
and $\lambda_{\pm}=\frac{t'_{z}\pm\sqrt{t'{}_{z}^{2}+\left(m'-2t_{z}\right)t_{z}}}{at_{z}}$.

Projecting the $xy$ part of the Hamiltonian (Eq. (\ref{eq:hxy in critical point})),
onto the subspace spanned by the two end modes (Eq. (\ref{eq:explicit form of end modes of h1d})),
we find a single anisotropic Dirac cone on the surface. The corresponding
Hamiltonian takes the form
\begin{equation}
\mathcal{H}_{xy\text{-surface}}=2a\left[k_{y}t_{y}\sigma_{y}+k_{x}t_{x}\sin\left(k_{\text{so}}a\right)\sigma_{x}\right].\label{eq:xy surface Dirac Hamiltonian}
\end{equation}

Two remarks are in order: first, one can derive the topological nature
of the model from the bulk wavefunctions. In fact, recognizing that
the system has an inversion symmetry, generated by the operator $\sigma{}_{x}\tau_{x}$,
we can implement the procedure introduced in Ref. \cite{Fu2007a} \textcolor{black}{(relating the $\mathbb{Z}_{2}$ invariant to the
parity of the occupied states at the time-reversal invariant points)}
and easily evaluate the $\mathbb{Z}_{2}$ invariant. We have verified
that the results of this analysis are consistent with the above derivation,
where the surface was studied directly.

Second, we note that since the system is guaranteed to preserve its
topological nature as long the gap remains finite, the various parameters
are not restricted to the values given above. In particular, the strict
requirement $t_{y}=t_{y}'$ can be relaxed.

In the next section we study the fractional analog of the model introduced in this section.

\subsection{Fractional strong topological insulators from weakly coupled wires}

\label{sub:fractional 3D}

In Sec. \ref{sub:review of 2D}, when constructing the Laughlin-like
$\nu=1/m$ state in 2D, we saw that one can define the $\eta$-fields,
in terms of which the problem is mapped to the simple $\nu=1$ case.
Reversing the logic, we see that by taking a topological state with
$\nu=1$, and replacing the $\phi$ fields by the $\eta$ fields,
we expect to get a fractional state.

In this Section, we follow this approach in generalizing our strong
topological insulator to its fractional analog. To do so, we start
by writing the Hamiltonian of the $\nu=1$ case, discussed in the
previous section, in terms of the bosonic $\phi$-fields. Then we
write the same Hamiltonian with the $\eta$-fields, and tune the system
to filling $\nu=1/m$, where such terms conserve momentum.

We note that all the fields which are not around $k_{x}=0$ are trivially
gapped by the intra-layer terms, as seen most clearly in Fig. (\ref{fig:diagram_integer}).
Consequently, the topological properties involve only the modes around
$k_{x}=0$. In what follows, we therefore write only the operators
acting on these fields. In addition, we omit the indices $R/L$, which
are fully determined by the spin indices for the $k_{x}=0$ modes.

We define the vector
\begin{align}
\tilde{\Psi}_{i,n} & =\left(\begin{array}[t]{cccc}
e{}^{i\eta_{1,\uparrow}^{i,n}}e & ^{i\eta_{2,\uparrow}^{i,n}} & e^{i\eta_{3,\uparrow}^{i,n}} & e^{i\eta_{4,\uparrow}^{i,n}}\end{array}\right.\label{eq:Capital psi tilde}\\
 & \hphantom{jijijjjj}\left.\begin{array}[t]{cccc}
e^{i\eta_{1,\downarrow}^{i,n}} & e^{i\eta_{2,\downarrow}^{i,n}} & e^{i\eta_{3,\downarrow}^{i,n}} & e^{i\eta_{4,\downarrow}^{i,n}}\end{array}\right)^{T},\nonumber
\end{align}
where $i$ represents the index enumerating the unit cells in each
layer, and $n$ counts the layers. In these notations, the low energy
Hamiltonian takes the form

\begin{widetext}

\begin{equation}
H_{x}=\sum_{n,i,\alpha,s}\frac{v}{2\pi m}\int dx\left(\partial_{x}\eta_{\alpha,s}^{i,n}\right)^{2},\label{eq:hx in terms of bosonization-1}
\end{equation}
\begin{equation}
H_{y}=-t_{y}\sum_{n,i}\int dx\left[\tilde{\Psi}_{i,n}^{\dagger}\left(\tau_{x}+\frac{1}{2}\left(\tau_{y}\sigma_{y}+\tau_{x}\sigma_{x}\right)\right)\tilde{\Psi}_{i,n}+\left(\frac{1}{4}\tilde{\Psi}_{i+1,n}^{\dagger}\left(\tau_{x}+i\tau_{y}\right)\left(\sigma_{x}+i\sigma_{y}\right)\tilde{\Psi}_{i,n}+h.c.\right)\right],\label{eq:hy in terms of bosonization}
\end{equation}

\begin{align}
H_{z} & =\frac{1}{2}\sum_{n,i}\int dx\left[\left(\tilde{\Psi}_{i,n+1}^{\dagger}\left(-t_{z}\tau_{x}+it'_{z}s_{y}\tau_{z}\right)\left(1-\sigma_{x}\right)\tilde{\Psi}_{i,n}+h.c\right)+\tilde{\Psi}_{i,n}^{\dagger}m'\tau_{x}\left(1-\sigma_{x}\right)\tilde{\Psi}_{i,n}\right].\label{eq:3D Hamiltonian in terms of bosonization}
\end{align}

\end{widetext} For simplicity, we do not consider the effects of
density-density interactions between the various modes.

We emphasize that analyzing the problem directly in terms of the bosons
is essential in the fractional case. Unfortunately, the bosonic form
makes it difficult to see that the above set of non-commuting terms
results in a gapped system with a gapless surface. In the $\nu=1$
case, we can of course refermionize the above Hamiltonian and repeat
the analysis of Sec. \ref{sub:integer 3D} to show this explicitly.
In the fractional case, where refermionization does not result in
a solvable model, the situation is more subtle as the various inter-wire
terms are irrelevant in the weak coupling limit. To avoid additional
complications that arise from that, we work in the regime where the
bare amplitudes are large enough such that the inter-wire terms flow to the strong coupling limit and successfully gap out the bulk.

Indeed, if the Hamiltonians $H_{y}$ and $H_{z}$
have very large bare amplitudes, we can neglect the intra-wire terms
and the physics becomes practically independent of $m$ because the
inter-wire Hamiltonian is quadratic in the fermionic $\tilde{\psi}$-fields.
Since we know from the fermionic language that the system is gapped
in the $m=1$ case, it is clear that in this limit the same is true
for $m>1$. This result is expected to remain true for moderately
large bare amplitudes as well.

\begin{figure}
\includegraphics[scale=0.5]{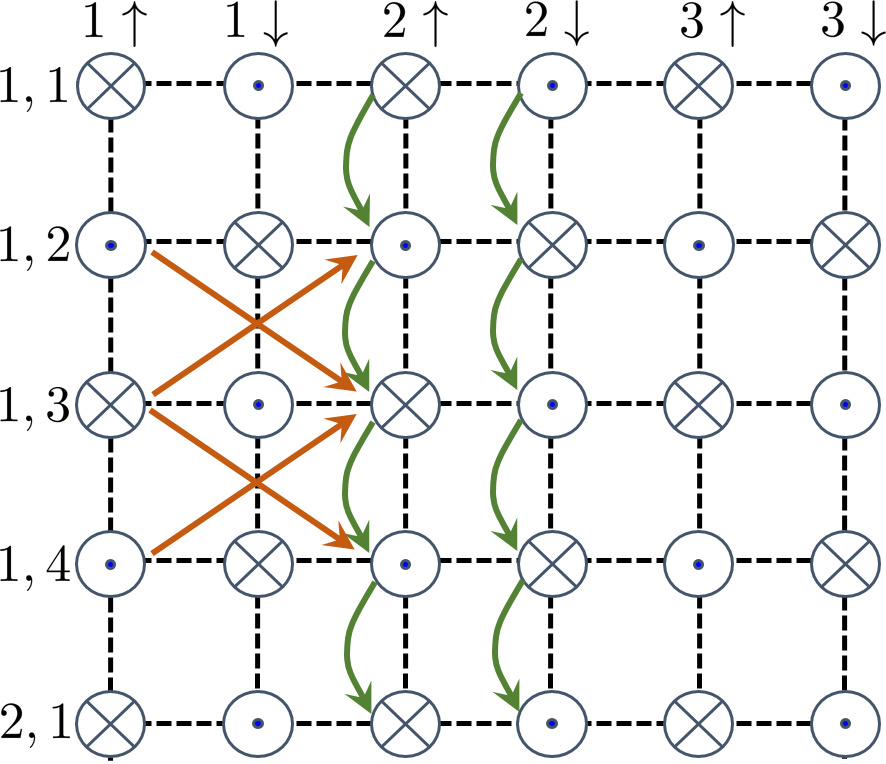}\protect\caption{\label{fig:reduce tight binding model}A diagrammatic representation
of the reduced 2D tight binding model, describing the inter-wire part of the full 3D model. Here,
each ``lattice point'' corresponds to a linearly dispersing mode
in the $x$ direction, with a well defined $y$ and $z$ coordinates.
The horizontal axis describes the layer and spin indices ($n,s$),
and the vertical axis describes the position within each layer ($i,\alpha$).
Again, the symbol $\bigotimes$ ($\bigodot)$ represents a right (left)
mover. The green arrows represent the terms in the intra-layer Hamiltonian
$H_{y}$, and the brown\textbf{ }arrows represent inter-layer terms
which couple the spin-up modes, and take the form $-2t_{z}\cos\left(k_{z}a\right)\tau_{x}$
in $k$-space. We emphasize that the reduced model
defined in Eq. (\ref{eq:hy in terms of bosonization})-(\ref{eq:3D Hamiltonian in terms of bosonization}) contains additional terms
which are not depicted here. If all the terms are considered, the full 3D model is topologically non-trivial
if the reduced model forms a 2D topological insulator. Thus, the analysis
of some aspects of the strongly interacting 3D phase is reduced to the
analysis of the 2D non-interacting topological phase.}
\end{figure}

In what follows, the topological nature of the Hamiltonian is revealed again
with the aid of diagrammatic representations. We depict the $\eta$
modes which are not gapped by intra-layer terms (i.e., the modes near
$k_{x}=0$) by diagrams of the form shown in Fig. (\ref{fig:reduce tight binding model}).
 As before, the symbol $\bigotimes$ ($\bigodot)$ represents a right
(left) mover, and colored arrows connecting two modes represent coupling
between them. The horizontal axis represents the layer and spin indices
($n,s$), and the vertical axis represents the intra-layer position
($i,\alpha$)\textbf{.}

Forgetting for a moment that the symbols represent dispersing 1D modes,
if they are treated as states localized on the corresponding lattice
points, the diagram presents the 3D problem as a 2D tight-binding
model, defined in real space by Eqs. (\ref{eq:hy in terms of bosonization})-(\ref{eq:3D Hamiltonian in terms of bosonization}).
The 2D lattice model describing the inter-wire coupling terms is referred
to as the reduced 2D model. The usefulness of the reduced tight binding
model description is revealed by noting that the full 3D model is
topologically non-trivial if the reduced model forms a
2D topological insulator.

If the system is infinite (or periodic) in the $\hat{y}$ and $\hat{z}$
directions, the corresponding momenta are good quantum numbers.
We therefore write the problem in terms of its Fourier components
$k_{y}$ and $k_{z}$. We note that in order to take advantage of
the bosonization description, we do not perform a Fourier transform
in the $x$ direction. Consequently, the Hamiltonian takes the form
\begin{equation}
H=H_{x}+\sum_{\mathbf{k}}\int dx\tilde{\Psi}_{\mathbf{k}}^{\dagger}(x)\left(h_{y}\left(\mathbf{k}\right)+h_{z}\left(\mathbf{k}\right)\right)\tilde{\Psi}_{\mathbf{k}}(x),\label{eq:Hamiltonian in bosnization}
\end{equation}
with $\mathbf{k}=\left(k_{y},k_{z}\right)$, and
\begin{equation}
\tilde{\Psi}_{\mathbf{k}}(x)=\frac{1}{\sqrt{N_{y}N_{z}}}\sum_{j,n}e^{-i\left(4jk_{y}a+nk_{z}a\right)}\tilde{\Psi}_{j,n}(x),\label{eq:fourier transform}
\end{equation}
\begin{align}
h_{y}\left(\mathbf{k}\right) & =-t_{y}\tau_{x}-\frac{t_{y}}{2}\left(\tau_{y}\sigma_{y}+\tau_{x}\sigma_{x}\right)\label{eq:h_xy (k)}\\
 & -\frac{t_{y}}{2}\left(\tau_{x}\sigma_{x}-\tau_{y}\sigma_{y}\right)\cos\left(4k_{y}a\right)\nonumber \\
 & -\frac{t_{y}}{2}\left(\tau_{x}\sigma_{y}+\tau_{y}\sigma_{x}\right)\sin\left(4k_{y}a\right),\nonumber
\end{align}
\begin{align}
h_{z}\left(\mathbf{k}\right) & =\frac{1}{2}\left[\left(m'-2t_{z}\cos\left(k_{z}a\right)\right)\tau_{x}\right.\label{eq:h_z (k)}\\
 & \hphantom{gfgg}+\left.2t'_{z}\sin\left(k_{z}a\right)s_{y}\tau_{z}\right]\left(1-\sigma_{x}\right),
\end{align}
 where $N_{y}$ is the number of wires in each layer, and $N_{z}$
is the number of layers (cf. Eqs. (\ref{eq:hxy}) and (\ref{eq:hz})).

If the 3D model is finite in the $\hat{y}$ or $\hat{z}$ direction,
the reduced 2D model has an edge. For $m'<2t_{z}$, it can be checked
that the reduced model forms a 2D topological insulator, and has counter
propagating edge modes. Recalling that each lattice point in the reduced
model represents a linearly dispersing $\eta$-mode, we argue that
these edge states correspond to the gapless surface modes of the original
3D model. Therefore, we can get the surface modes by diagonalizing
the reduced tight binding model and finding the corresponding edge
states.

In order to get an explicit analytic form, we focus again on the regime
where $2t_{z}$ is close to $m'$, and the gap becomes small. If the
system terminates at the $z=0$ plane, the reduced model has
two edge modes which take the form
\begin{equation}
\xi_{\beta}(x,k_{y})=\frac{1}{\sqrt{N_{y}}}\sum_{j,n}\sqrt{a}f(a\cdot n)e^{-ijk_{y}a}A_{\beta}\cdot\tilde{\Psi}_{j,n}(x),\label{eq:capital phi}
\end{equation}
where $\beta=1,2$ labels the two counter-propagating edge modes,
$f(z)$ is the function defined after Eq. (\ref{eq:explicit form of end modes of h1d}),
and the vectors $A_{\beta}$ are
\begin{align}
A_{1} & =\frac{1}{4}\left(\begin{array}[t]{cccc}
1+i & ,1-i & ,-i-1 & ,i-1\end{array}\right.,\nonumber \\
 & \left.\begin{array}[t]{cccc}
\phantom{,}\phantom{}\hphantom{}\hphantom{}-i+1 & ,-i-1 & ,-1+i & ,1+i\end{array}\right)\nonumber \\
A_{2} & =A_{1}^{*}.\label{eq:A}
\end{align}

In the integer case, where $m=1$, Eq. (\ref{eq:capital phi}) can
be thought of as a bosonic description of the Dirac mode localized
near the $z=0$ surface (we point out
the similarity between the fields defined in Eqs. (\ref{eq:capital phi})
and (\ref{eq:explicit form of end modes of h1d}), up to a different
choice of basis).
In the fractional case, where $m>1$, the above gapless modes cannot
be described by Dirac's theory of free fermions. As before, we refer to these
more general modes as fractional Dirac modes.

The part of the inter-wire Hamiltonian describing the edge modes of
the reduced model (which correspond to the surface mode of the original
3D model) takes the form
\begin{align}
H_{edge} & =2t_{y}a\sum_{k_{y}}\int dxk_{y}\vec{\xi}^{\dagger}(x,k_{y})\bar{\tau}_{z}\vec{\xi}(x,k_{y})\nonumber \\
 & =-2t_{y}i\int dxdy\vec{\xi}^{\dagger}(x,y)\bar{\tau}_{z}\partial_{y}\vec{\xi}(x,y),\label{eq:H_edge}
\end{align}
where $\vec{\xi}=\left(\begin{array}[t]{cc}
\xi_{1} & \xi_{2}\end{array}\right)^{T}$, and $\bar{\tau}_{z}$ is a Pauli matrix acting on this basis.
To simplify the analysis presented in the next section, the second
line of Eq. (\ref{eq:H_edge}) was written in the continuum limit.

In the next section we will use the reduced 2D model formalism to
study the properties of the surface once it is gapped by breaking
its protecting symmetries. We will see that the resulting gapped fractional
Dirac modes have unique properties which distinguish them from massive
Dirac fermions.

\subsection{Gapping the surface}

\label{sub:Gapping the surface}

\subsubsection{Breaking time-reversal symmetry: Halved  fractional
quantum Hall effect on the surface}

\label{sub:fractionaltheta}

Having written a low energy effective surface Hamiltonian using the
reduced model, we turn to study what happens when the surface is gapped
by breaking time reversal symmetry.

Like in a non-interacting strong topological insulator, we expect the effective
action describing the response of our system to the electromagnetic
field to contain a $\theta$-term of the form $\frac{\theta}{2\pi}\frac{e^{2}}{2\pi}\int d^{3}xdt\vec{E}\cdot\vec{B}$.
In general, time reversal symmetry severely restricts the possible
values that the axion angle $\theta$ can take. Naively, one would
expect $\theta$ to be either 0, corresponding to a trivial insulator,
or $\pi$, corresponding to a topological insulator \cite{Qi2008}.
However, it was shown in Ref. \cite{Maciejko2010} that fractional
axion angles may be consistent with time-reversal symmetry in topologically
ordered systems. As we will see, our fractional strong topological
insulator provides an example of such a scenario.

In general, the $\theta$-term in the effective action implies that
breaking time-reversal symmetry on the surface results in a non-zero
surface Hall conductance of the form $\sigma_{xy}=\pm\frac{\theta}{2\pi}\frac{e^{2}}{h}$.
In what follows, we calculate $\sigma_{xy}$ directly and then use
the above general result to determine the axion angle $\theta$ characterizing
our system. To do so, we first break time-reversal symmetry by introducing
a Zeeman field $B$ on the surface. We examine a configuration where
$B$ changes sign as we cross the line $y=0$. By studying the properties
of the gapless mode attached to the boundary, we will be able to deduce
the surface Hall conductance.

Within the reduced 2D model formalism, the problem of finding the
1D channel attached to a magnetic domain wall on the surface is converted
into that of finding the localized zero-energy mode associated with
a similar domain wall on the edge of a 2D topological insulator. To
be concrete, we use the continuum model described by Eq. (\ref{eq:H_edge})
and add a space dependent perturbation of the form $B(y)\vec{\xi}^{\dagger}\bar{\tau}_{x}\vec{\xi}$,
where $B(y)=B_{0}\text{sign}\left(y\right)$. This Hamiltonian, and
hence the full inter-wire Hamiltonian, has a zero-energy solution
of the form
\begin{equation}
\xi_{B}(x)=\sqrt{\lambda_{B}}a\sum_{j,n}f\left(n\cdot a\right)e^{-\lambda_{B}\left|j\right|a}\beta\cdot\tilde{\Psi}_{j,n}(x),\label{eq:Fractional fermion mode}
\end{equation}
with
\begin{equation}
\beta=\frac{1}{2}\left(\begin{array}[t]{cccccccc}
0 & 1 & 0 & -1 & 1 & 0 & -1 & 0\end{array}\right),\label{eq:B}
\end{equation}
and $\lambda_{B}=\frac{B_{0}}{2t_{y}a}$.

Notice that this mode, being a combination of right movers, is a right
moving mode. Furthermore, we argue that it is identical to the chiral
mode that resides on the edge of a $\nu=1/m$ Laughlin QHE state.

To see this, we calculate the electron propagator characterizing this
1D channel, defined as $G_{B}(x-x',t-t')=\left\langle \tilde{\xi}_{B}(x',t')\tilde{\xi}_{B}^{\dagger}(x,t)\right\rangle $.
Recall that $\xi_{B}(x)$ is an exact zero-energy solution of the
Inter-wire Hamiltonian for any $x$. Therefore, neglecting the coupling
to the gapped excitations of the inter-wire Hamiltonian, we can calculate
the expectation value with respect to the quadratic intra-wire Hamiltonian
shown in Eq. (\ref{eq:hx in terms of bosonization-1}) . This results
in the propagator
\begin{equation}
G_{B}(x,t)\propto\frac{1}{\left(x-vt\right)^{m}},\label{eq:propagator}
\end{equation}
which is indeed identical the one characterizing the edge of a $\nu=1/m$
QHE edge state \cite{Wen2007}.

Invoking the bulk edge correspondence, we find that
$\sigma_{xy}=\pm\frac{1}{2m}\frac{e^{2}}{h}$, in agreement with Sec.
\ref{sub:surface of strong}. The halved fractional surface quantum
Hall effect indicates that the system is characterized by a fractional
axion angle $\theta=\pi/m$. A direct implication of the above is
that the system has a fractional magneto-electric response.

We note that to get to Eq. (\ref{eq:propagator})
we have used the full intra-wire Hamiltonian, which includes contributions
from modes which are already gapped by the inter-wire terms. In doing
so, we implicitly assume that the intra-wire Hamiltonian does not
couple $\tilde{\xi}_{B}$ to these gapped fields. This can be justified
by calculating $\left\langle \left(\tilde{\xi}'\right)^{\dagger}H_{x}\tilde{\xi}_{B}\right\rangle ,$
where $\tilde{\xi}'$ is one of the gapped modes and the average is
taken with respect to the intra-wire Hamiltonian. Using the definitions
of $\tilde{\xi_{B}}$ and $H_{x}$ {[}Eqs. (\ref{eq:hx in terms of bosonization-1})
and (\ref{eq:Fractional fermion mode}){]} and the fact that $\tilde{\xi}'$
must be orthogonal to $\tilde{\xi_{B}}$, a straightforward calculation
shows that this average indeed vanishes, justifying the above calculation.

\subsubsection{Coupling the surface to a superconductor: The emergence
of a fractional Majorana mode}

\label{sub:fractionalMajorana}

Next, we turn to ask what happens when the surface is gapped by breaking
charge conservation. This is done by proximity coupling the surface
to an $s$-wave superconductor. It was found in Ref. \cite{Fu2008}
that when the surface of a strong topological insulator is gapped
in this fashion, the resulting phase resembles a spinless $p_{x}+ip_{y}$
superconductor, but has time-reversal symmetry.

In our model, this problem corresponds to understanding what happens
to the edge of the reduced 2D topological insulator when it is coupled
to an $s$-wave superconductor. This motivates us to write a proximity
term of the form $\Delta\left(\xi_{1}^{\dagger}\xi_{2}^{\dagger}+h.c.\right)$,
coupling the two edge modes of the reduced model. Using the enlarged
basis, $\vec{\xi}_{N}=\left(\begin{array}[t]{cccc}
\xi_{1}(x,y) & \xi_{2}(x,y) & \xi_{2}^{\dagger}(x,y) & \xi_{1}^{\dagger}(x,y)\end{array}\right)^{T}$, we rewrite the edge Hamiltonian in the form
\begin{equation}
\frac{1}{2}\int dxdy\vec{\xi}_{N}^{\dagger}\left(-2it_{y}\tau_{z}\sigma_{z}\partial_{y}+\frac{\Delta}{a}\tau_{z}\sigma_{x}\right)\vec{\xi}_{N}.\label{eq:Edge hamiltonian with delta}
\end{equation}

In order to reveal the topological nature of the superconducting phase,
we consider the boundary between a region with a non-zero $B$ a region
with a non-zero $\Delta$. Studying non-interacting strong topological
insulators, the authors in Ref. \cite{Fu2008} have found that such
a boundary contains a chiral Majorana mode. We now examine what happens
in the fractional case.

In the presence of time-reversal breaking and superconducting terms,
the edge part of the reduced Hamiltonian takes the form
\begin{align}
 & \frac{1}{2}\int dxdy\vec{\xi}_{N}^{\dagger}\left[-2it_{y}\tau_{z}\sigma_{z}\partial_{y}\right.\label{eq:Edge Hamiltonian with delta and B}\\
 & \left.+\frac{B(y)}{a}\tau_{x}\sigma_{z}+\frac{\Delta(y)}{a}\tau_{z}\sigma_{x}\right]\vec{\xi}_{N}.\nonumber
\end{align}
We consider a simple situation, where $B(y)=B_{0}\Theta(y)$ and
$\Delta=\Delta_{0}\Theta(-y)$ (to be concrete, we assume $\Delta_{0}>0,B_{0}>0$).

The Hamiltonian given by Eq. (\ref{eq:Edge Hamiltonian with delta and B})
has a zero energy solution of the form
\begin{equation}
\frac{1}{\sqrt{2}}\left(\xi_{B\Delta}+\xi_{B\Delta}^{\dagger}\right),\label{eq:fractional majorana mode}
\end{equation}
with
\begin{equation}
\xi_{B\Delta}=\sqrt{\frac{2\lambda_{B}\lambda_{\Delta}}{\lambda_{B}+\lambda_{\Delta}}}a\sum_{j,n}e^{-i\pi/4}f(n\cdot a)\beta\cdot\tilde{\Psi}_{j,n}(x)g(ja),\label{eq:gamma_B delta}
\end{equation}
\begin{equation}
g(y)=\left\{ \begin{array}{c}
e^{-\lambda_{B}y},y>0\\
e^{\lambda_{\Delta}y},y<0
\end{array}\right.,\label{eq:g(y)}
\end{equation}
 $\lambda_{\Delta}=\frac{\Delta_{0}}{2t_{y}a}$ and $\lambda_{B}=\frac{B_{0}}{2t_{y}a}$.\textbf{
}For $m=1$, we find a self-Hermitian combination of right moving
fermions, making it a chiral Majorana mode, in agreement with Ref.
\cite{Fu2008}. If $m>1$, we find again a self-Hermitian chiral mode.
However, in this case it cannot be described
by a free Majorana theory. In particular, the tunneling density of
states associated with this mode is proportional to $\omega^{m-1}$,
as opposed to the free Majorana case, where the density of states
is constant. This is in fact similar to the tunneling density of states
characterizing the edge of a $\nu=1/m$ Laughlin QHE states. To find
this result, one can repeat the process that led to Eq. (\ref{eq:propagator})
and write the propagator of the above fractional Majorana mode. The
above results are in agreement with Sec. \ref{sub:surface of strong},
where we have used a modified time-reversal symmetry to directly model
the surface.

\section{Discussion}

The effects of strong interactions on topological
matter is a complex subject, and a very active field of study. The
challenge becomes even larger in 3D due to the limited analytical
tools, and the technical difficulties in applying the existing numerical
methods. The coupled-wires approach has been shown to be useful in
the theoretical study of such phases in 2D. Within this approach,
the existing machinery for treating 1D interacting
systems is used to produce various analytical results, which reflect
the topological nature of the phase in question.

In this paper, we have demonstrated that the coupled-wires
approach can be used to model and analyze 3D topological systems.
We have focused on the fractional counterpart of the well known strong
topological insulator, which demonstrates the remarkable physical
properties of 3D topologically ordered phases. While it is not trivial
to study the bulk excitations and the gapless surface in the current
formulation, we were able to analyze the non-trivial characteristics
of the surface once it is gapped. In particular, the coupled wires
approach has been shown to be very effective in revealing the nature
of 1D modes residing in the vicinity of domain walls between
distinct gapped regions.

This allowed us to show that if the
surface is gapped by breaking time-reversal symmetry, it is characterized
by a halved fractional Hall conductivity of the form $\sigma_{xy}=\frac{1}{2m}\frac{e^{2}}{h}$.
Furthermore, if the surface is partitioned into a magnetic region
and a superconducting region, a novel fractional Majorana mode was
shown to emerge on the boundary between them.

It would additionally be interesting to generalize the configuration
discussed in Ref. \cite{Akhmerov2009}, in which one can study and
electrically detect the interferometry of Majorana fermions on the
surface of a strong topological insulator, to our fractional case.
Such an electronic Mach-Zehnder interferometer is expected to contain
signatures that differentiate the fractional Majorana modes from free
Majorana modes. This will be elaborated on elsewhere.

The current approach provides a path for
constructing the fractional analogues of non-interacting topological
phases. It would be interesting to systematically apply this formulation
to the different classes found in the periodic table of topological
phases \cite{Schnyder2008,Kitaev2009}, as was done for 2D systems
in Ref. \cite{Neupert2014}. The current work provides the necessary
tools for extensions to 3D phases.

{\it Note added.} We point out a recent related work (Ref. \cite{Meng2015b}), in which an extension of the coupled-wires approach to 3D was discussed.  We note that the phases constructed in the above paper are composed of quantum Hall subsystems, and are therefore fundamentally different from the 3D fractional phases constructed in the current work.

\begin{acknowledgments}
We are indebted to Jason Alicea, Erez Berg, Iliya Esin, Liang Fu, Arbel Haim, David Mross, Xiao-Liang Qi, Raul Santos, Eran Sela, and Ady Stern
 for insightful conversations. We acknowledge the support
of the Israel Science Foundation (ISF), the Minerva Foundation, and
the European Research Council under the European Community's Seventh
Framework Program (FP7/2007-2013)/ERC Grant agreement No. 340210.
ES is supported by the Adams Fellowship Program of the Israel Academy
of Sciences and Humanities.
\end{acknowledgments}

\bibliographystyle{apsrev4-1}
%

\end{document}